\documentclass{article}

\pdfoutput=1 
\usepackage{jheppub}
\usepackage{graphicx,verbatim}
\usepackage{psfrag, times}
\usepackage[T1]{fontenc}
\usepackage{epsf}
\usepackage{mathtools}
\usepackage[dvipsnames]{xcolor}

\usepackage{amsmath}
\usepackage{amsfonts}
\usepackage{amssymb}
\usepackage{mathrsfs}
\usepackage{slashed}
\usepackage{bbm}

\def\be{\begin{equation}}
\def\ee{\end{equation}}
\def\bc{\begin{center}}
\def\ec{\end{center}}

\def\bR{{\mathbb{R}}}

\def\bbC{{\mathbb{C}}}

\def\bbX{{\mathbb{X}}}

\def\bbR{{\mathbb{R}}}

\def\bN{{\bf{N}}}
\def\bK{{\bf{K}}}

\def\cH{{\mathcal{H}}}

\def\cM{{\mathcal{M}}}
\def\cN{{\mathcal{N}}}

\def\cU{{\mathcal{U}}}

\def\cW{{\mathcal{W}}}

\def\cZ{{\mathcal{Z}}}

\def\r2{{\sqrt{2}}}
\def\h{{\eta}}

\def\bea{\begin{eqnarray}}
\def\eea{\end{eqnarray}}

\def\bK{{\bf{K}}}

\def\bc{{\bf{c}}}

\def\bbZ{{\mathbb{Z}}}
\def\bN{{\bf{N}}}

\def\rx{{\rm{x}}}

\def\fq{{\mathfrak{q}}}

\def\bi{{\mathbf{i}}}
\def\bj{{\mathbf{j}}}
\def\bl{{\mathbf{l}}}

\def\eps{\epsilon}
\def\cH{{\mathcal{H}}}

\usepackage{url}
\usepackage[mathscr]{euscript}

\usepackage{dynkin-diagrams}
\title{Quantum Integrable Systems from Supergroup Gauge Theories}
\author[]{Heng-Yu Chen${}^{1}$, Taro Kimura${}^{2}$ and Norton Lee${}^{3}$}
\affiliation{$^1$\rm Department of Physics, National Taiwan University, Taipei 10617, Taiwan}
\affiliation{$^2$\rm Institut de Math\'ematiques de Bourgogne, Universit\'e Bourgogne Franche-Comt\'e, 21078 Dijon, France}
\affiliation{$^3$\rm C. N. Yang Institute for Theoretical Physics, Stony Brook University, Stony Brook, NY 11794, USA}
\emailAdd{heng.yu.chen@phys.ntu.edu.tw}
\emailAdd{taro.kimura@u-bourgogne.fr}
\emailAdd{norton.lee@stonybrook.edu}
\vspace{2cm}
\abstract{In this note, we establish several interesting connections between the supergroup gauge theories and the super integrable systems, i.e. gauge theories with supergroups as their gauge groups and integrable systems defined on superalgebras. In particular, we construct the super-characteristic polynomials of super-Toda lattice and elliptic double Calogero-Moser system by considering certain orbifolded instanton partition functions of their corresponding supergroup gauge theories.
We also derive an exotic generalization of $\mathfrak{sl}(2)$ XXX spin chain arising from the instanton partition function of SQCD with supergauge group, and study its Bethe ansatz equation.}

\begin{document}

\maketitle

\newpage
\section{Introduction and Summary}

It is well-known that supergroups and their underlying superalgebras are the natural extensions of Lie groups and Lie algebras by including Grassmannian valued generators. In the studies of supersymmetric field theories, 
these supergroups typically encode both bosonic and fermionic global space time symmetries. 
However when considered as the local or gauge symmetries, a quantum field theory with a supergroup as its gauge group is inevitably non-unitary due to the violation of spin-statistics, therefore this class of theories has often been overlooked and not been explored much in the literature. 
In more recent years, thank to the various advancements in computational techniques, there have been renewed interests in them and broaden our general understanding in these exotic quantum field theories.
Most notably, the explicit constructions of supergroup gauge via D-branes and so-called negative/ghost D-branes were given in \cite{Okuda:2006fb, Dijkgraaf:2016lym,Nekrasov:2018xsb} (See \cite{Vafa:2001qf} also for earlier work), moreover the  partition functions for supergroup gauge instantons have been computed in \cite{Kimura:2019msw} using equivariant localization. We will review both of these developments momentarily.

%\paragraph{}
One of the active themes in the studies of supersymmetric gauge theories in various dimensions has been uncovering their underlying mathematical structures and often connecting them with classical and quantum integrable systems. This research theme
often leads us to obtain useful quantities for given integrable systems as computed by gauge theoretical means. For example, the generating function of commuting conserved Hamiltonians known as \textit{characteristic polynomial}, can be identified with the generating function of gauge theory chiral ring \cite{Martinec:1995by, Gorsky:1994dj}. We can further identify the spectral curves of integrable models to the Seiberg-Witten (SW) curve of gauge theory. Well-known examples include $\cN=2$ SYM/Toda lattice~\cite{Martinec:1995by} and $\cN=2^*$/Calogero-Moser (CM) system~\cite{Donagi:1995cf,DHoker:1997hut}. 
%\paragraph{}

It is natural to investigate whether these well-established exact correspondences between the gauge theories and integrable systems can be extended to their respective super versions. 
In this paper, we will initiate the study on such an interesting correspondence between various supersymmetric integrable systems and supergroup gauge theories. 
Moreover at the classical level, 
we can naturally generalize the Lax operators/matrices of ordinary integrable systems to
construct so-called super-Lax operators which take values on the supermatrices. Several super versions of the known integrable systems have been constructed in this way, including Super-Toda lattice \cite{van1994super}, super spin chain \cite{Evans:1996bu,Beisert:2003yb}, and double Calogero-Moser (dCM) system, a super analog of the well known Calogero-Moser (CM) system, both trigonometric (tdCM)~\cite{HiJack, sergeev2001superanalogs} and elliptic double Calogero-Moser systems (edCM)~\cite{Chen:2019vvt, Nekrasov:2017gzb}. 
Integrability, which is again defined by the existence of many commuting conserved Hamiltonians, of dCM system was proved in \cite{HiJack} for the trigonometric, and in \cite{Chen:2019vvt} for the elliptic cases.
%\paragraph{}

At the quantum level, we would also like to extend the earlier identifications of the commuting quantum Hamiltonians with the quantum chiral ring operators \cite{Nikita-Shatashvili, HYC:2011, Dorey:2011pa} to the supergroup cases. To achieve this, in addition to putting the supersymmetric gauge theories on the so-called $\Omega$-background and taking appropriate limit \cite{Nekrasov:2012xe},
it is important recognize the need for further introducing the co-dimension two surface defect in the gauge theory through appropriate orbifolding procedures \cite{Kanno:2011fw,Nakajima:2011yq}. 
Introducing such a surface defect (more precisely full surface defect) into the gauge theory generates chainsaw quiver structure.
For $G$-gauge theory, such a quiver structure is equipped with a set of gauge couplings $(\fq_i)_{i = 1,\ldots,\operatorname{rk}G}$, which will later be identified with the canonical coordinates of the underlying quantum integrable systems.
This allows us to reproduce the quantum conserved Hamiltonians from gauge theory~\cite{Chen:2019vvt}, and the further identification between instanton partition function and quantum wave function \cite{Nekrasov:2017gzb}. 
We will generalize these procedures to the various supergroup gauge  theories.

In the subsequent sections,
we will first review on D-brane realization \cite{Dijkgraaf:2016lym} of supergroup gauge theories and their instanton configurations \cite{Kimura:2019msw} in Section~\ref{supergroup Gauge}. We then reconstruct the conserved Hamiltonians from the instanton configuration of supergroup gauge theory following the procedure in \cite{Chen:2019vvt}: We first show that the saddle point equations of super instanton partition function under Nekrasov-Shatashvili (NS) limit reproduces Bethe ansatz equation (BAE) of the corresponding super integrable system, and further calculate the conserved Hamiltonians of the integrable systems. We have successfully established gauge/integrability correspondence for pure supergroup gauge theory and super-Toda lattice in Section~\ref{super-Toda}, supergroup QCD/super-XXX spin chain in Section~\ref{super-XXX}, and $\cN=2^*$/edCM system in Section~\ref{edCM}.

\section{Supergroup Gauge Theory}\label{supergroup Gauge}

\subsection{D-brane Construction}
%\paragraph{}

One important application of string theory is the construction of various gauge theories using D-branes. For the gauge theories with ordinary Lie groups as their gauge groups, the D-brane constructions are well-known. The D-brane constructions of the gauge theories with supergroups as their gauge groups were first considered in~\cite{Vafa:2001qf,Okuda:2006fb}, and later studied in various other papers such as \cite{Kapustin:2009cd,Witten:2011zz,Dijkgraaf:2016lym,Mikhaylov:2014aoa,Mikhaylov:2015nsa,Okazaki:2017sbc,Aghaei:2018cbn}. 
To construct a D-brane system realizing super gauge groups, one need to introduce the concept of \textit{negative brane} (or \textit{ghost brane}) for which we will have a brief review below, see~\cite{Okuda:2006fb, Dijkgraaf:2016lym} for details. 
%\paragraph{}

On the string world sheet, D-brane acts as its boundary. In the presence of multiple D-branes, a Chan-Paton factor, {which takes a value in the vector space $\mathbb{C}^N$ for a stack of $N$ D-branes,} is  assigned to each string in order to identify which D-brane it ends on. These Chan-Paton factors, though non-dynamical on the world sheet, give rise to $N$ labels that generates $U(N)$ gauge group in the target space. 
The Dirac quantization requires $N$ to be an integer, this however does not exclude the possibility of $N$ being negative. A negative Chan-Paton factor can be realized by introducing additional minus sign to certain subset of boundary states, i.~e. the Chan-Paton factor now takes the value in a graded vector space $\mathbb{C}^{n_+|n_-}$ with the mixed signatures:
\begin{equation}\label{metric}
     \begin{pmatrix}
     +\mathbf{1}_{n_+} & 0 \\
    0 & -\mathbf{1}_{n_-}
    \end{pmatrix}.
\end{equation}
The additional vector space with negative signature corresponds to the additional $n_-$ \textit{negative branes} which carry an additional negative sign with respect to the boundary states hence the Chan-Paton factors corresponding to the usual \textit{positive branes}. This distinguishes them from the more commonly encountered anti-branes which only carry an negative sign for the the RR sector, and preserve the opposite space-time supersymmetries. A system consisting of both positive and negative branes can preserve the same space time supersymmetries. Moreover due to the negative sign in their NS-NS sector, a negative brane carries negative tension and anti-gravitates, this feature again differs from the anti-brane which gravitates.
%\paragraph{}

Let us focus on the world volume theory of a stack of coinciding $n_+$ positive D3 and $n_-$ negative D3 branes.
The extra minus sign introduced in negative brane means an open string connecting a positive and a negative branes have the opposite of usual statistics. An open string with both ends on negative branes have usual statistics since the two minus signs from both ends cancel. This enhances an $U(n_+)\times U(n_-)$ quiver gauge group to an $U(n_+|n_-)$ super-gauge group, i.e. a supermatrix unitary group, {$U \in \operatorname{End}(\mathbb{C}^{n_+|n_-})$ with $U^{-1} = U^\dag$},
\begin{align}
    U=
    \begin{pmatrix}
    A & B \\
    C & D
    \end{pmatrix}
    \in U(n_+|n_-)
\end{align}
where $A$ and $D$ are $n_+\times n_+$ and $n_- \times n_-$ even graded matrices with complex number entries and $B$ and $C$ are $n_+\times n_-$ and $n_-\times n_+$ odd graded matrices whose entries are Grassmannian  \eqref{super odd}. 
An odd graded matrix means its elements are subject to anti-commutation relation instead of commutation relation, \begin{align}
    B_{ij}C_{kl}=(-1)C_{kl}B_{ij}.
\end{align}

Let us consider an $\cN=4$ super Yang-Mills (SYM) theory with super gauge group $U(n_+|n_-)$ realized in the world volumes of $n_+$ positive D3-branes and $n_-$ negative D3-branes, the Lagrangian density can be written as:
\begin{align}\label{Lagrangian}
    \frac{1}{g^2}\left[\operatorname{str}F^2+\sum_{i=1}^6\operatorname{str}(D\Phi^i)^2+\cdots\right]=\frac{1}{g^2}\left[\operatorname{tr}F^2_+-\operatorname{tr}F_-^2+\sum_{i=1}^6\operatorname{tr}(D\Phi^i_+)^2-\sum_{i=1}^6\operatorname{tr}(D\Phi^i_-)^2+\cdots\right],
\end{align}
where we have only listed out the contributions charged under the bosonic subgroup $U(n_+)\times U(n_-)\subset U(n_+|n_-)$, and we denote them with subscripts $+$ and $-$. 
There are however few obvious features which distinguish the gauge theory with $U(n_+|n_-)$  and $U(n_+)\times U(n_-)$ gauge group.
First we notice from \eqref{Lagrangian}
that the kinetic energy is unbounded from below thus the $U(n_+|n_-)$ theory is manifestly non-unitary. 
This arises from the relative minus sign in \eqref{Lagrangian} between positive and negative field strength gauge coupling $\displaystyle \tau_+=-\tau_-:=\tau=\frac{4\pi i}{g^2}+\frac{\theta}{2\pi}$. 
The Yang-Mills action can be evaluated in the instanton background as \cite{Kimura:2019msw, osborn1981semiclassical}:
\begin{align}
    \exp\left[-\frac{2\pi i\tau}{16\pi^2}\int \operatorname{str}F^2\right]=\exp[2\pi i\tau (k_+-k_-)]=e^{2\pi i\tau_+k_+}e^{2\pi i\tau_-k_-},
\end{align}
where $k_\pm$ are both non-negative integers.
As the result, the instanton configurations of supergroup consist of positive and negative instantons, with the counting parameters  $\fq_\pm=e^{2\pi i\tau_\pm}$ now satisfy the relation: $\fq_+=\fq_-^{-1}:=\fq=e^{2\pi i\tau}$.
In addition, we will see momentarily in the localization computation that two bosonic subgroups of $U(n_+|n_-)$ are oppositely charged under $\Omega$-background deformation, this is different from the  $U(n_+)\times U(n_-)$ quiver gauge theory whose two sub-groups are equally charged under $\Omega$-background deformation.

\subsection{Instanton Moduli Space for Supergroup Gauge Theory}\label{adhm}
%\paragraph{}

We will start with a brief review about the construction of supergroup instanton configuration, namely Atiyah-Drinfeld-Hitchin-Manin (ADHM) construction~\cite{Atiyah:1978ri} of super instantons, more details can be found in \cite{Kimura:2019msw}.
Let us define the graded vector spaces:
\begin{subequations}\label{vecsp}
\begin{align}
    \bN&=\bbC^{n_+|n_-}=\bbC^{n_+}\oplus\bbC^{n_-}=\bN_+\oplus \bN_-\label{vecspN}\\
    \bK&=\bbC^{k_+|k_-}=\bbC^{k_+}\oplus\bbC^{k_-}=\bK_+\oplus\bK_-\label{vecspK}
\end{align}
\end{subequations}
where the $\pm$-sign in the subscript indicates the signature of the respective sub-space.  
{These are the Chan-Paton factors of the corresponding brane configuration:}
\begin{center}
\begin{tabular}{|c|c|c|c|c|c|c|c|c|c|c|c|c|}
	\hline
	Brane Type & Signature of Chan-Paton factor & \# of Branes & 1 & 2& 3& 4& 5& 6& 7& 8& 9& 10 \\ \hline\hline
	$\text{D}(-1)_{+}$ & Positive & $k_+$ & & & & & & & & & & \\ \hline
	$\text{D}(-1)_{-}$ & Negative & $k_-$ & & & & & & & & & & \\ \hline
	$\text{D3}_{+}$ & Positive & $n_{+}$ & x& x& x& x& & & & & & \\ \hline
	$\text{D3}_{-}$ & Negative & $n_{-}$ & x& x& x& x& & & & & & \\ \hline
\end{tabular}
\end{center}
The ADHM matrices can be realized as the open strings stretching between the D(-1)-D3 configuration.
The matrices $B_{1,2}\in\text{Hom}(\bK,\bK)$ label the open strings having both ends on D(-1) branes, $I\in\text{Hom}(\bN,\bK)$ and $J\in\text{Hom}(\bK,\bN)$ represents the open strings with one end attached to D(-1) branes and the other end on D3 branes. In particular $I$ and $J$ can be further decomposed into:
\begin{align}
    I=I_+\oplus I_-;\quad J=J_+\oplus J_-
\end{align}
where $I_\sigma\in\text{Hom}(\bN_\sigma,\bK)$, $J_\sigma\in\text{Hom}(\bK,\bN_\sigma)$ for $\sigma=\pm$.
Defining the instanton moduli space by
\begin{align}\label{moduli space}
    \cM_{n_+,n_-,k_+,k_-}=\{(B_1,B_2,I,J) \mid \mu_\bR=\zeta_{>0}\, {\rm diag} (1_{{\bf K}_+},-1_{{\bf K}_-}),\ \mu_\bbC=0\}/U(k_+|k_-),
\end{align}
where the real and complex  momentum maps are:
\begin{subequations}
\begin{align}
    \mu_\bR&=[B_1,B_1^\dagger]+[B_2,B_2^\dagger]+II^\dagger-J^\dagger J \\
    % =[B_1,B_1^\dagger]+[B_2,B_2^\dagger]+I_+I_+^\dagger-I_-I_-^\dagger-J_+^\dagger J_++J_-^\dagger J_-; \\
    \mu_\bbC&=[B_1,B_2]+IJ
    % =[B_1,B_2]+I_+J_+-I_-J_-.
\end{align}
\end{subequations}
We can write real momentum map in terms of supermatrices (in the form \eqref{supermatrix blockform}) 
\begin{align}
    \mu_\bR = &  
    \begin{pmatrix}
    [B_{1,00},B_{1,00}^\dagger] + [B_{1,01},B_{1,01}^\dagger] & B_{1,00}B_{1,10}^\dagger + B_{1,01}B_{1,11}^\dagger - B_{1,10}B_{1,00}^\dagger - B_{1,11}B_{1,01}^\dagger \\
    B_{1,10}B_{1,00}^\dagger + B_{1,11}B_{1,01}^\dagger - B_{1,00}B_{1,10}^\dagger - B_{1,01}B_{1,11}^\dagger & [B_{1,10},B_{1,10}^\dagger] + [B_{1,11},B_{1,11}^\dagger]
    \end{pmatrix} \nonumber\\
    & + \begin{pmatrix}
    [B_{2,00},B_{2,00}^\dagger] + [B_{2,01},B_{2,01}^\dagger] & B_{2,00}B_{2,10}^\dagger + B_{2,01}B_{2,11}^\dagger - B_{2,10}B_{2,00}^\dagger - B_{2,11}B_{2,01}^\dagger \\
    B_{2,10}B_{2,00}^\dagger + B_{2,11}B_{2,01}^\dagger - B_{2,00}B_{2,10}^\dagger - B_{2,01}B_{2,11}^\dagger & [B_{2,10},B_{2,10}^\dagger] + [B_{2,11},B_{2,11}^\dagger]
    \end{pmatrix} \nonumber\\
    & + 
    \begin{pmatrix}
    I_{00}I_{00}^\dagger + I_{01}I_{01}^\dagger & I_{00}I_{10}^\dagger + I_{01}I_{11}^\dagger \\
    I_{10}I_{00}^\dagger + I_{11}I_{01}^\dagger & I_{10}I_{01}^\dagger + I_{11}I_{11}^\dagger
    \end{pmatrix} - 
    \begin{pmatrix}
    J_{00}^\dagger J_{00} + J_{10}^\dagger J_{10} & J_{00}^\dagger J_{01} + J_{10}^\dagger J_{11} \\ 
    J_{01}^\dagger J_{00} + J_{11}^\dagger J_{10} & J_{01}^\dagger J_{01} + J_{11}^\dagger J_{11}
    \end{pmatrix} 
\end{align}
We can do the same for $\mu_\bbC$:
\begin{align}
    \mu_{\bbC} = &
    \begin{pmatrix}
    B_{1,00}B_{2,00} + B_{1,01}B_{2,10} & B_{1,00}B_{2,01} + B_{1,01}B_{2,11} \\
    B_{1,10}B_{2,00} + B_{1,11}B_{2,10} & B_{1,10}B_{2,01} + B_{1,11}B_{2,11}
    \end{pmatrix} +
    \begin{pmatrix}
    I_{00}J_{00} + I_{01}J_{10} & I_{00}J_{01} + I_{01}J_{11} \\
    I_{10}J_{00} + I_{11}J_{10} & I_{10}J_{01} + I_{11}J_{11}
    \end{pmatrix} 
\end{align}
Moduli space \eqref{moduli space} has 2 types of solutions based on property of parameter $\zeta$ (up to $U(k_+|k_-)$ transformation)
\begin{itemize}
    \item If $\zeta$ is a real number, the moduli space is bosonic, i.e. $B_1$, $B_2$, $I$, and $J$ are diagonal.
    \item If $\zeta$ is grassmannian even, the moduli space is fermionic, i.e. $B_1$, $B_2$, $I$, and $J$ have only off-diagonal component.
\end{itemize}
Here we choose $\zeta$ as a real positive number to serve our purpose.
We will now show that the instanton vector space $\bK=\bK_+\oplus\bK_-$ is generated by two independent sets of ADHM data by proving the following stability condition:
\begin{align}
    \bK_+=\bbC[B_1,B_2]I_+(\bN_+);\quad \bK_-= \bbC[B_1^\dagger,B_2^\dagger]J^\dagger_-(\bN_-).
\end{align}
\textbf{Proof: }
We will prove this condition by contradiction.
Let us denote
\begin{align}
    \bK_+'=\bbC[B_1,B_2]I_+(\bN_+)\subset\bK_+;\quad \bK_-'= \bbC[B_1^\dagger,B_2^\dagger]J^\dagger_-(\bN_-)\subset\bK_-.
\end{align}
Assuming there exists  $\bK^\perp_\pm=\bK_\pm-\bK_\pm'$, 
we can define the projection operators
$P_\pm:\bK\to\bK^\perp_\pm$, $P_\pm=P_\pm^2=P_\pm^\dagger$ such that $P_\pm B_{1,2}=B_{1,2}P_\pm$, $P_\pm I_+=P_\pm J_-^\dagger=0$. The projection only acts on one of the graded sub-spaces $P_+I_-=0$, $P_-J_+^\dagger=0$. We will prove this by contradiction that existence of such $\bK^\perp_\pm$ violates the stability condition defining moduli space in \eqref{moduli space}.
One may consider the projection of real momentum map $\mu_\bbR$ to $\bK^\perp_+$ space:
\begin{align}
    \zeta_{>0}\mathbf{1}_{\bK^\perp_+}=P_+\mu_\bR P_+=[(P_+B_1P_+),(P_+B_1P_+)^\dagger]+[(P_+B_2P_+),(P_+B_2P_+)^\dagger]-P_+J_+J_+^\dagger P_+.
\end{align}
Taking the trace on the both sides of the equation above yields:
\begin{align}
    0\leq\text{tr}(\zeta_{>0}\mathbf{1}_{\bK_+^\perp})=-\text{tr}(P_+J^\dagger_+J_+P_+)\leq0
\end{align}
This means $\bK^\perp_+=0$. A similar argument applies to $P_-$, and one finds that $\bK_-^\perp=0$ also. \textbf{q.e.d.}
%\paragraph{}

By using the stability condition, the moduli space can now be written via Hyper K\"{a}hler quotient:
\begin{align}
    {\cM}_{n_+,n_-,k_+,k_-}=\{(B_1,B_2,I,J)\mid\mu_\bbC=0\}/\!\!/\operatorname{GL}(k_+|k_-).
\end{align}
The action of local transformation $g\in\text{GL}(k_+|k_-)$ and $h\in\text{GL}(n_+|n_-)$ on ADHM variables is given by
\begin{align}\label{local}
    (g,h):(B_1,B_2,I,J)\to(gB_1g^{-1},g B_2 g^{-1},gIh^{-1},hJg^{-1}).
\end{align}
The instanton partition function of supergroup gauge theory is equivalent integral over ADHM moduli space:
\begin{align}
    \mathcal{Z}^{\rm inst}_{U(n_+|n_-)} = \sum_{k_+,k_-} \fq^{k_+-k_-}\mathcal{Z}_{U(n_+|n_-)}^{k_+,k_-} \quad \text{with} \quad \cZ_{U(n_+|n_-)}^{k_+,k_-} = \int_{{\cM}_{n_+,n_-,k_+,k_-}}1.
    \label{eq:instpart}
\end{align}
We will evaluate \eqref{eq:instpart} by following standard Nekrasov theory. Let us introduce the $\Omega$-deformation 
\begin{align}
    (q_1,q_2):(B_1,B_2,I,J)\to (q_1B_1,q_2B_2,I,q_1q_2J),
\end{align}
such that the fixed points on the moduli space ${\cM}_{n_+,n_-,k_+,k_-}$ are defined where the $\Omega$-deformation can be canceled by local transformation \eqref{local}. The volume integration over \eqref{eq:instpart} is regularized to counting of discrete fixed points on the moduli space.
Using the stability condition, one can show that $I_-=0=J_+$ at the fixed points. This implies that $B_1$ and $B_2$ commute at the fixed points. 
Each saddle point in $\bK_\pm$ can be labeled by
\begin{align}
    \label{eq:ADHMsp}
    \bK_+: \, B_1^{i-1}B_2^{j-1}I_+(\bN_+) , \quad \bK_-: \, (B_1^{\dagger})^{i-1}(B_2^\dagger)^{j-1}J_-^\dagger(\bN_-).
\end{align}
This means that each fixed point can be labeled by a set of partitions (Young diagrams) $(\vec{\lambda}^+,\vec{\lambda}^-)=\{\lambda^+_{\alpha=1,\dots,n_+}, \lambda^-_{\beta=1,\dots,n_-} \}$.
Since our system is defined upon supergroup, the super-character of ADHM data is taken by supertrace: 
\begin{subequations}
\begin{align}
    N=\operatorname{sch}(\bN)&=\operatorname{ch}(\bN_+)-\operatorname{ch}(\bN_-)=N_+-N_-, \\
    K=\operatorname{sch}(\bK)&=\operatorname{ch}(\bK_+)-\operatorname{ch}(\bK_-)=K_+-K_-.
\end{align}
\end{subequations}
The characters can be written as:
\begin{subequations}\label{ADHM character}
\begin{align}
    &N_+=\sum_{\alpha=1}^{n_+} e^{a^+_\alpha},\qquad K_+=\sum_{\alpha=1}^{n_+} e^{a_\alpha^+}\sum_{(i,j)\in\lambda_\alpha^+}q_1^{i-1}q_2^{j-1},  \\
    &N_-=\sum_{\beta=1}^{n_-} e^{a_\beta^-}, \qquad K_-=\sum_{\beta=1}^{n_-} e^{a_\beta^-}q_+^{-1}\sum_{(i,j)\in\lambda_\beta^-}q_1^{1-i}q_2^{1-i},
\end{align}
\end{subequations}
where $q_+=q_1q_2$. The additional power of $q_+^{-1}$ comes from the fact that $J$ carries a charge of $q_1q_2=q_+$ in the $\Omega$-background.
%\paragraph{}

For each positive and negative instanton, we have instanton counting parameter $\fq_\pm=e^{2\pi i\tau_\pm}$. As stated, the bosonic moduli space \eqref{moduli space} is defined based on diagonal ADHM data.
By similar manipulation of the localization calculation producing standard Nekrasov partition function \cite{Kimura:2019msw,Nekrasov:2002qd,Nekrasov:2003rj}, each ADHM data and equations contributes to the Euler character class of moduli space tangent bundle by \cite{Nekrasov:2015wsu,Kimura:2019msw}: 
\begin{itemize}
    \item ADHM matrix $B_1$ contributes to tangent space character by $q_1KK^*$,
    \item AdHM matrix $B_2$ contributes to tangent space character by $q_2KK^*$,
    \item ADHN matrix $I$ contributes to tangent space character by $NK^*$,
    \item ADHM matrix $J$ contributes to tangent space character by $q_{1}q_{2}N^*K$,
    \item Local $U(k_+|k_-)$ transformation on vector space ${\bf K}$ contributes to character by $-KK^*$,
    \item Equation $\mu_\bbC=0$ contributes to character by $-q_{1}q_{2}KK^*$.
\end{itemize}
we can construct the tangent space character
\begin{align}
    \mathcal{T}_{\vec{\lambda}_+,\vec{\lambda}_-} = NK^* + q_{12}N^*K - (1-q_1)(1-q_2)KK^* 
\end{align}
with $q_i = e^{\epsilon_i}$. We may now denote instanton partition function as sum over Young diagrams:
\begin{align}\label{Z-pure}
    \cZ^{\text{inst}}_{U(n_+|n_-)}=\sum_{\vec{\lambda}^+,\vec{\lambda}^-}\fq^{|\vec{\lambda}^+|-|\vec{\lambda}^-|}\cZ_{U(n_+|n_-)}[\vec{\lambda}^+,\vec{\lambda}^-]; \quad \cZ_{U(n_+|n_-)}[\vec{\lambda}^+,\vec{\lambda}^-]=\mathbb{E}\left[-\frac{SS^*}{P_{12}^*}+\frac{NN^*}{P_{12}^*}\right],
\end{align}
where the dual bundle of $X$ is denoted by $X^*$, and we define the universal bundle characters
\begin{align}
    S = S_+ - S_- , \qquad
    S_\pm=N_\pm-P_1P_2K_\pm,
    \label{UnivBundle}
\end{align}
with
\begin{align}
    P_i=1-q_i , \qquad
    P_{12} = (1 - q_1)(1 - q_2),
\end{align}
which should not be confused with the projector $P_\pm$ mentioned before.
The operator $\mathbb{E}$ converts the additive Chern characters to the multiplicative classes:
\begin{align}\label{E operator}
    \mathbb{E}\left[ \sum_i \sigma_i e^{x_i}  \right]
    = \prod_{i} x_i^{\sigma_i}
\end{align}
where $\sigma_i = \pm 1$ is the sign factor associated with each Chern root.
This is also obtained as the cohomological limit of the plethystic exponential. 

The measure in \eqref{eq:instpart} can be expressed into the following contour integration \cite{Kimura:2019msw} by denoting character $K = \sum_{i=1}^{k_+}e^{\phi_i^+} - \sum_{j=1}^{k_-}e^{\phi_j^-}$:
\begin{align}\label{Super-Contour-Z}
    \cZ_{U(n_+|n_-)}^{k_+,k_-} = 
    \oint_{\Gamma_+\times \Gamma_-} & \prod_{\sigma=\pm}\left[\prod_{s=1}^{k_\sigma} \frac{d\phi_a^\sigma}{2\pi i}z_{n_\sigma,k_{\sigma}}^{\text{inst}}(\phi^\sigma,a^\sigma)\right] \prod_{s=1}^{k_+} \prod_{t=1}^{k_-} \frac{(\phi_{st}^{+-2}-\eps_1^2)(\phi_{st}^{+-2}-\eps_2^2)}{\phi_{st}^{+-2}(\phi_{st}^{+-2}-\eps_+^2)} \nonumber\\
    & \times \prod_{s=1}^{k_+}\prod_{\beta=1}^{n_-}(\phi_s^+-a_\beta^-)(\phi_s^+-a_\beta^- + \epsilon_+) \prod_{t=1}^{k_-}\prod_{\alpha=1}^{n_+}(\phi_t^- - a_{\alpha}^+ ) (\phi_t^- - a_\alpha^+ + \epsilon_+) 
\end{align}
where $\phi_{st}^{\sigma\sigma'}=\phi_s^{\sigma} - \phi_t^{\sigma'}$ and with
\begin{align}
    z_{n,k}^{\text{inst}}(\phi,a) = \frac{1}{k!}\left(\frac{\eps_+}{\eps_1\eps_2}\right)^k \prod_{s<t}^k \frac{\phi_{st}^2(\phi_{st}^2-\eps_+^2)}{(\phi_{st}^2-\eps_1^2)(\phi_{st}^2-\eps_2^2)} \prod_{s=1}^k \prod_{\alpha=1}^n \frac{1}{(\phi_s-a_\alpha)(\phi_s-a_\alpha+\eps_+)}.
    \label{eq:z_inst}
\end{align}
Let us assume that $\epsilon_1$, $\epsilon_2$ have positive imaginary part. The contour $\Gamma_\pm$ is chosen to be
\begin{subequations}
\begin{align}
    &\Gamma_+ = (\mathbb{R}+ i a_{\rm Im}^+ )^{k_+}, \quad a_{\rm Im}^+ = \underset{{\alpha=1,\dots,n_+}}{\min}\{{\rm Im} (a_{\alpha}^+)\} \\
    &\Gamma_- = (\mathbb{R} + i a_{\rm Im}^{-})^{k_-}, \quad a_{\rm Im}^- = \underset{\beta=1,\dots,n_-}{\max}\{{\rm Im} (a_{\beta}^-)\}.
\end{align}
\end{subequations}
Notice that the power counting of integration variable $\phi^\sigma_s$ is of degree -2, which vanishes quickly when integration variables approaching infinity, we can therefore enclose our contour accordingly. 
The contour integration can now be evaluated by familiar residue formula, and the poles picked up by integration variables $\phi^{\sigma}$ are 
\begin{subequations}
\begin{align}
    \phi^+: \quad & a_{\alpha}^+ + (i-1)\epsilon_1 + (j-1)\epsilon_2, \quad \alpha = 1,\dots,n_+, \ i,j\in\mathbb{N} \\
    \phi^-: \quad & a_{\beta}^- - i\epsilon_1 - j\epsilon_2, \quad \beta=1,\dots,n_-, \ i,j\in\mathbb{N}
\end{align}
\end{subequations}
These poles correspond to the fixed point condition for the ADHM variables~\eqref{eq:ADHMsp} under the equivariant action.
\paragraph{}
More generally, if we consider introducing the D0-branes into the ${\rm D4}_{\pm}$-NS5 brane configuration which engineers the five-dimensional quiver super-gauge groups in \cite{Kimura:2019msw}, \cite{Dijkgraaf:2016lym}, they will play the role of the gauge instantons.
When we dimension reduce this combined D-brane configuration, 
we can thus interpret the super-instanton partition function \eqref{Super-Contour-Z} as integrating over the zero-mode fluctuations around the moduli space of the $\rm{D(-1)}_{\pm}$/D-instanton world volume (super) matrix model.

\section{Super Toda Lattice and Pure $\cN=2$ Supergroup Gauge Theory}\label{super-Toda}
%\paragraph{}

The gauge/integrability correspondence can be calculated in various aspects \cite{Nikita-Shatashvili,Nekrasov:2015wsu,Nekrasov:2017gzb,Chen:2019vvt}. One important development is the identification of q-character~\cite{Frenkel:1998,Knight:1995JA} (or called T-function) of gauge theory with the characteristic polynomial of the associated quantum integrable model. For ordinary group, we are able to extract every conserved Hamiltonian of the integrable model from q-character, see \cite{Chen:2019vvt} for details. 
We will show that the same procedure can be extended to supergroup case. In particular in this chapter we will reconstruct Hamiltonian of super Toda lattice from pure $\cN=2$ supergroup gauge theory.

\subsection{Bethe Ansatz Equation from Instanton Partition Function}

Starting with the instanton partition function of pure $\cN=2$ SYM theory with $U(n_+|n_-)$ gauge group is given in~\eqref{Z-pure}, which consists of four pieces
\begin{equation}
\begin{aligned}\label{Z-nm pure}
\mathcal{Z}_{U(n_+|n_-)}^\text{inst}
& =\sum_{(\vec{\lambda}^+,\vec{\lambda}^-)}\fq_+^{|\vec{\lambda}^+|}\fq_-^{|\vec{\lambda}^-|}\mathbb{E}\left[-\frac{SS^*}{P^*_{12}}+\frac{NN^*}{P_{12}^*}\right], \\
& =\sum_{(\vec{\lambda}^+,\vec{\lambda}^-)}\fq^{|\vec{\lambda}^+|-|\vec{\lambda}^-|}\mathcal{Z}^{++}_\text{vec}\mathcal{Z}^{-+}_\text{vec}\mathcal{Z}^{+-}_\text{vec}\mathcal{Z}^{--}_\text{vec}, \\
& =\sum_{(\vec{\lambda}^+,\vec{\lambda}^-)}\fq^{|\vec{\lambda}^+|-|\vec{\lambda}^-|}\prod_{\sigma,\sigma'=\pm }\mathcal{Z}^{\sigma\sigma'}_\text{vec}.
\end{aligned}
\end{equation}
In this section, we would like to extend the methods in \cite{Chen:2019vvt} to supergauge group, the main goal is to extract conserved commuting Hamiltonians from the $T$-function or q-character which we will construct subsequently. 

The fixed points on the instanton moduli space are labeled by a set of Young diagrams:
\begin{align}
    \{\vec{\lambda}^+,\vec{\lambda}^-\}=\{(\lambda^{+,(1)},\dots,\lambda^{+,(n_+)}),(\lambda^{-,(1)},\dots,\lambda^{-,(n_-)})\}
\end{align}
where each Young diagram is represented by a row vector $\lambda^{\sigma,(\alpha)}=(\lambda^{\sigma,(\alpha)}_1,\lambda^{\sigma,(\alpha)}_2,\dots,)$ with non-negative integer entries labeling the number of boxes such that:
\begin{align}
    \lambda^{\sigma,(\alpha)}_{i}\geq\lambda^{\sigma,(\alpha)}_{i+1}.
\end{align} 
We define the following parameters 
\begin{subequations}\label{x data}
\begin{align}
    &x_{\alpha i}^+=a_\alpha^++(i-1)\eps_1+\epsilon_2\lambda^{+,(\alpha)}_{i}; \qquad \mathring{x}^{+}_{\alpha i}=a_\alpha+(i-1)\eps_1, \\
    &x_{\beta i}^-=a_\beta^--i\eps_1-\epsilon_2\lambda^{-,(\beta)}_{i}; \qquad \mathring{x}^{-}_{\beta i}=a_\beta^--i\eps_1,
\end{align}
\end{subequations}
such that
\begin{subequations}\label{S+-}
\begin{align}
    S_+&=N_+-P_{12}K_+=\sum_{\alpha=1}^{n_+}e^{a^+_\alpha}+P_1\sum_{\alpha=1}^{n_+}\sum_{i=1}^\infty \left( e^{x_{\alpha i}^{+}}-e^{\mathring{x}^{+}_{\alpha i}} \right) :=P_1X_+;\label{S+} \\
    S_-&=N_--P_{12}K_-=\sum_{\beta=1}^{n_-}e^{a_\beta^-}-P_1\sum_{\beta=1}^{n_-}\sum_{i=1}^\infty \left( e^{x^-_{\beta i}}-e^{\mathring{x}^{-}_{\beta i}} \right) :=-P_1X_-,\label{S-}
\end{align}
\end{subequations}
with 
\begin{align}
    X_+=\sum_{\alpha=1}^{n_+}\sum_{i=1}^\infty e^{x^+_{\alpha i}},\quad X_-=\sum_{\beta=1}^{n_-}\sum_{i=1}^\infty e^{x^-_{\beta i}}.
\end{align}
By its definition, $\cZ^{++}_{\text{vec}}$ is exactly the same as ordinary gauge group which had been considered in various earlier papers \cite{HYC:2011,Nikita-Shatashvili,Chen:2019vvt}, thus we will not repeat the calculations here but only give the final result:
\begin{align}\label{Z++}
    \cZ_\text{vec}^{++}[\lambda^+]&=\mathbb{E}\left[-\frac{S_+S_+^*}{P_{12}^*}+\frac{N_+N_+^*}{P_{12}^*}\right]=\left.\prod_{\alpha=1}^{n_-}\prod_{\beta=1}^{n_-}\prod_{i=1}^\infty\prod_{j=1}^\infty\right|_{(\alpha i)\neq(\beta j)} \frac{\Gamma(\epsilon_2^{-1}(x_{\alpha i}^+-x_{\beta j}^+-\epsilon_1))}{\Gamma(\epsilon_2^{-1}(x_{\alpha i}^+-x_{\beta j}^+))}\cdot\frac{\Gamma(\epsilon_2^{-1}(\mathring{x}^{+}_{\alpha i}-\mathring{x}_{\beta j}^{+}))}{\Gamma(\epsilon_2^{-1}(\mathring{x}^{+}_{\alpha i}-\mathring{x}_{\beta j}^{+}-\epsilon_1))}.
\end{align}
All the remaining $\cZ_\text{vec}^{\sigma\sigma'}$ in \eqref{Z-nm pure} are newly introduced by the supergroup structure. We will first present the results and then discuss how each term is obtained: 
\begin{subequations}
\begin{align}
    \cZ_\text{vec}^{--}[\vec{\lambda}^-]&=\left.\prod_{\alpha=1}^{n_-}\prod_{\beta=1}^{n_-}\prod_{i=1}^\infty\prod_{j=1}^\infty\right|_{(\alpha i)\neq(\beta j)}\frac{\Gamma(\epsilon_2^{-1}(x^{-}_{\alpha i}-x^{-}_{\beta j}-\epsilon_1))}{\Gamma(\epsilon_2^{-1}(x^{-}_{\alpha i}-x^{-}_{\beta j}))}
    \frac{\Gamma(\epsilon_2^{-1}(\mathring{x}^{-}_{\alpha i}-\mathring{x}^{-}_{\beta i}))}{\Gamma(\epsilon_2^{-1}(\mathring{x}^{-}_{\alpha i}-\mathring{x}^{-}_{\beta j}-\epsilon_1))}\label{Z--}; \\
    \cZ^{+-}_\text{vec}[\vec{\lambda}^+,\vec{\lambda}^-]&=\prod_{\alpha=1}^{n_+}\prod_{\beta=1}^{n_-}\prod_{i=1}^\infty\prod_{j=1}^\infty\frac{\Gamma(\epsilon_2^{-1}(x^{+}_{\alpha i}-x^{-}_{\beta j}-\epsilon_1))}{\Gamma(\epsilon_2^{-1}(x^{+}_{\alpha i}-x^{-}_{\beta j}))}
    \frac{\Gamma(\epsilon_2^{-1}(\mathring{x}^{+}_{\alpha i}-\mathring{x}^{-}_{\beta i}))}{\Gamma(\epsilon_2^{-1}(\mathring{x}^{+}_{\alpha i}-\mathring{x}^{-}_{\beta j}-\epsilon_1))} \label{Z+-};\\
    \cZ^{-+}_\text{vec}[\vec{\lambda}^+,\vec{\lambda}^-]&=\prod_{\alpha=1}^{n_-}\prod_{\beta=1}^{n_+}\prod_{i=1}^\infty\prod_{j=1}^\infty\frac{\Gamma(\epsilon_2^{-1}(x^{-}_{\alpha i}-x^{+}_{\beta j}-\epsilon_1))}{\Gamma(\epsilon_2^{-1}(x^{-}_{\alpha i}-x^{+}_{\beta j}))}
    \frac{\Gamma(\epsilon_2^{-1}(\mathring{x}^{-}_{\alpha i}-\mathring{x}^{+}_{\beta i}))}{\Gamma(\epsilon_2^{-1}(\mathring{x}^{-}_{\alpha i}-\mathring{x}^{+}_{\beta j}-\epsilon_1))} \label{Z-+}.
\end{align}
\end{subequations}
Here $\cZ^{--}_\text{vec}$ is instanton partition function of $U(n_-)$ gauge group by definition. We may simply replace $\{[x^+], [\mathring{x}^{+}]\}$ in \eqref{Z++} by $\{[x^-],[\mathring{x}^{-}]\}$, which one obtains~\eqref{Z--}. The direct supporting character calculation yields the following results:
\begin{align}
    \cZ^{--}_\text{vec}[\vec{\lambda}^-]&=\mathbb{E}\left[-\frac{S_-S_-^*}{P_{12}}+\frac{N_-N_-^*}{P_{12}^*}\right], \nonumber\\
    &=\mathbb{E}\left[-\frac{P_1X_-X_-^*}{P_2^*}+\frac{P_1\mathring{X}_{-}\mathring{X}_-^{*}}{P_2^*}\right], \nonumber\\
    &=\prod_{(\alpha i)\neq(\beta j)}\frac{\Gamma(\epsilon_2^{-1}(x^{-}_{\alpha i}-x^{-}_{\beta j}-\epsilon_1))}{\Gamma(\epsilon_2^{-1}(x^{-}_{\alpha i}-x^{-}_{\beta j}))}
    \frac{\Gamma(\epsilon_2^{-1}(\mathring{x}^{-}_{\alpha i}-\mathring{x}^{-}_{\beta i}))}{\Gamma(\epsilon_2^{-1}(\mathring{x}^{-}_{\alpha i}-\mathring{x}^{-}_{\beta j}-\epsilon_1))},
\end{align}
with $N_\pm=P_1\mathring{X}_\pm$.
To obtain the last line, we assume $|q_2|<1$ and use the following equation
\begin{align}
    \mathbb{E}\left[\frac{e^{-x}}{P_2^*}\right]=\mathbb{E}\left[\frac{e^x}{P_2}\right]=\mathbb{E}\left[\sum_{i=0}^\infty e^{x}q_2^{i}\right]=\prod_{i=0}^\infty\frac{1}{(x+i\epsilon_2)}=\Gamma(\eps_2^{-1}x)(\epsilon_2)^\frac{x}{\epsilon_2}.
\end{align}
The last equation is obtained from Gauss's product representation of Gamma function: 
\begin{align}
    \Gamma(z)=\lim_{m\to\infty}\frac{m^zm!}{z(z+1)\cdots(z+m)}.
\end{align}
The mixed contribution terms $\cZ^{+-}_\text{vec}[\vec{\lambda}^+,\vec{\lambda}^-]$ and $\cZ^{-+}_\text{vec}[\vec{\lambda}^+,\vec{\lambda}^-]$ can also be obtained from direct calculations:
\begin{subequations}
\begin{align}
    \cZ^{+-}_\text{vec}[\vec{\lambda}^+,\vec{\lambda}^-]
    &=\mathbb{E}\left[\frac{S_+S_-^*}{P_{12}^*}-\frac{N_+N_-^*}{P_{12}^*}\right], \nonumber\\
    &=\mathbb{E}\left[-\frac{P_1X_+X_-^*}{P_2^*}+\frac{P_1\mathring{X}_{+}\mathring{X}_{-}^*}{P_2^*}\right], \nonumber\\
    &=\prod_{(\alpha i),(\beta j)}\frac{\Gamma(\epsilon_2^{-1}(x^{+}_{\alpha i}-x^{-}_{\beta j}-\epsilon_1))}{\Gamma(\epsilon_2^{-1}(x^{+}_{\alpha i}-x^{-}_{\beta j}))}
    \frac{\Gamma(\epsilon_2^{-1}(\mathring{x}^{+}_{\alpha i}-\mathring{x}^{-}_{\beta i}))}{\Gamma(\epsilon_2^{-1}(\mathring{x}^{+}_{\alpha i}-\mathring{x}^{-}_{\beta j}-\epsilon_1))}, \\
    \cZ^{-+}_\text{vec}[\vec{\lambda}^+,\vec{\lambda}^-]&=
    \mathbb{E}\left[\frac{S_-S_+^*}{P_{12}^*}-\frac{N_-N_+^*}{P_{12}^*}\right], \nonumber\\
    &=\mathbb{E}\left[-\frac{P_1X_-X_+^*}{P_2^*}+\frac{P_1\mathring{X}_{-}\mathring{X}_+^{*}}{P_2^*}\right], \nonumber\\
    &= \prod_{(\alpha i),(\beta j)}\frac{\Gamma(\epsilon_2^{-1}(x^{-}_{\alpha i}-x^{+}_{\beta j}-\epsilon_1))}{\Gamma(\epsilon_2^{-1}(x^{-}_{\alpha i}-x^{+}_{\beta j}))}
    \frac{\Gamma(\epsilon_2^{-1}(\mathring{x}^{-}_{\alpha i}-\mathring{x}^{+}_{\beta i}))}{\Gamma(\epsilon_2^{-1}(\mathring{x}^{-}_{\alpha i}-\mathring{x}^{+}_{\beta j}-\epsilon_1))}.
\end{align}
\end{subequations}
%\paragraph{}

Next we would like to consider Nekrasov-Shatashvili limit \cite{Chen:2019vvt,HYC:2011} (NS-limit for short) such that $\epsilon_2\to0$ with $\epsilon_1=:\epsilon$ fixed \cite{Nikita-Shatashvili} for the various pieces we constructed above. In this limit, the vector multiplet contributions can be approximated by:
\begin{eqnarray}
\begin{aligned}
\mathcal{Z}^{\sigma\sigma'}_\text{vec}\approx\exp\left[\frac{1}{2\epsilon_2}\right.\sum_{(\alpha i)\neq(\alpha' i')}
&f(x^\sigma_{\alpha i}-x^{\sigma'}_{\alpha' i'}-\epsilon)-f(x^\sigma_{\alpha i}-x^{\sigma'}_{\alpha' i'}+\epsilon) \\
& \left.-f(\mathring{x}_{\alpha i}^{\sigma}-\mathring{x}_{\alpha' i'}^{\sigma'}-\epsilon)+f(\mathring{x}_{\alpha i}^{\sigma}-\mathring{x}_{\alpha' i'}^{\sigma'}+\epsilon)\right],
\end{aligned}
\end{eqnarray}
using Stirling approximation of $\Gamma$-function 
and we have introduced the function $f(x)=x(\log x-1)$. Moreover, the scaled variable $\epsilon_2\lambda_{i}^{\sigma,(\gamma)}$ becomes continuous, such that the sum over discrete Young diagrams can be approximated by continuous integral over a set of infinite integration variables
$\{x^+_{\alpha i},x^-_{\beta i}\}$ defined in \eqref{x data}.
To sum up we have:
\begin{equation}
\begin{aligned}
\mathcal{Z}_{U(n_+|n_-)}^{\text{inst}}
& =\int\prod_{\alpha i}dx^+_{\alpha i}\prod_{\beta i}dx^-_{\beta i}\exp\left[\frac{1}{\epsilon_2}\mathcal{H}_\text{inst}([x^+],[x^-])\right], \\
&=\int\prod_{\alpha i}dx^+_{\alpha i}\prod_{\beta i}dx^-_{\beta i}\exp\left[\frac{1}{\epsilon_2}\sum_{\sigma,\sigma'=\pm}\mathcal{H}^{\sigma\sigma'}([x^\sigma],[x^{\sigma'}])+\frac{1}{\epsilon_2}\sum_{\sigma=\pm}\mathcal{H}^\sigma([x^\sigma])\right],
\end{aligned}
\end{equation}
where the instanton function $\cH_\text{inst}([x^+],[x^-])$ are
\begin{equation}
\begin{aligned}
& \mathcal{H}^{\sigma\sigma'}([x^\sigma],[x^{\sigma'}])=\mathcal{U}^{\sigma\sigma'}([x^\sigma],[x^{\sigma'}])-\mathcal{U}^{\sigma\sigma'}([\mathring{x}^{\sigma}],[\mathring{x}^{\sigma'}]) \\
&\mathcal{H}^\sigma([x^{\sigma}])=\log\fq\sum_{(\alpha i)}x^{\sigma}_{\alpha i}-\log\fq\sum_{(\alpha i)}\mathring{x}^{\sigma}_{\alpha i}
\end{aligned}
\end{equation}
and
\begin{equation}\label{cU}
\begin{aligned}
\mathcal{U}^{\sigma\sigma'}=\frac{1}{2}\sum_{(\gamma i)\neq(\gamma' i')}
& 	f(x^\sigma_{\gamma i}-x^{\sigma'}_{\gamma' i'}-\epsilon)-f(x^\sigma_{\gamma i}-x^{\sigma'}_{\gamma' i'}+\epsilon).
\end{aligned}
\end{equation}
We will introduce the instanton density $\rho(x)$, which is unity along $\mathfrak{J}=\mathfrak{J}^+\cup\mathfrak{J}^-$, $\mathfrak{J}^+=\bigcup_{\alpha i}[\mathring{x}^{+}_{\alpha i},x^+_{\alpha i}]$, $\mathfrak{J}^-=\bigcup_{\beta i}[\mathring{x}^{-}_{\beta i},x^-_{\beta i}]$, and vanishes otherwise. Furthermore we also define
\begin{equation}
{G}(x)=\frac{d}{dx}\log\frac{(x-\epsilon)}{(x+\epsilon)};\qquad{R}^{\sigma'}(x)=\frac{1}{P^{\sigma'}(x)P^{\sigma'}(x+\epsilon)}
\end{equation}
where $P^+(x)=\prod_{\alpha=1}^{n_+}(x-a^+_\alpha)$ and $P^-(x)=\prod_{\beta=1}^{n_-}(x-a^-_\beta)$. 
Combining all together we may rewrite instanton partition function $\cH_\text{inst}$ as
\begin{equation}
\mathcal{H}^{\sigma\sigma'}=-\frac{1}{2}\operatorname{PV}\int_{\mathfrak{J}^\sigma}dx\int_{\mathfrak{J}^{\sigma'}}dy \, {G}(x-y)+\int_{\mathfrak{J}^\sigma}dx \log({R}^{\sigma'})
\end{equation}
where the symbol ``PV'' denotes the principle value integral. Under NS-limit, the whole integration on $\mathcal{H}_\text{inst}$ is dominated by saddle point configuration, the variation of $\rho(x)$ can be done by either varying $x^+_{\alpha i}$ or $x^-_{\beta i}$.
\begin{equation}
\begin{aligned}
0=\frac{\delta\mathcal{H}_\text{inst}}{\delta x^+_{\alpha i}}
&=\frac{\delta\mathcal{H}^{++}}{\delta x^+_{\alpha i}}+\frac{\delta\mathcal{H}^{+-}}{\delta x^+_{\alpha i}}+\frac{\delta\mathcal{H}^{-+}}{\delta x^+_{\alpha i}}+\frac{\delta\mathcal{H}^{+}}{\delta x^+_{\alpha i}}, \\
&=-\int_{\mathfrak{J}^+} dy \, {G}(x^+_{\alpha i}-y)+\int_{\mathfrak{J}^-}dy \, {G}(x^+_{\alpha i}-y)+\log({R}^+(x^+_{\alpha i}))-\log({R}^-(x^+_{\alpha i}))+\log\fq; \\
0=\frac{\delta\mathcal{H}_\text{inst}}{\delta x^-_{\beta i}}
&=\frac{\delta\mathcal{H}^{+-}}{\delta x^-_{\beta i}}+\frac{\delta\mathcal{H}^{-+}}{\delta x^-_{\beta i}}+\frac{\delta\mathcal{H}^{--}}{\delta x^-_{\beta i}}+\frac{\delta\mathcal{H}^{-}}{\delta x^-_{\beta i}} \\
&=+\int_{\mathfrak{J}^+} dy \, {G}(x^-_{\beta i}-y)-\int_{\mathfrak{J}^-}dy \, {G}(x^-_{\beta i}-y)-\log({R}^+(x^-_{\beta i}))+\log({R}^-(x^-_{\beta i}))-\log\fq.
\end{aligned}
\end{equation}
Since ${G}(x)$ is a total derivative, the two saddle point equations can be simplified to
\begin{equation}
\begin{aligned}
-1
& =\fq\frac{Q^+(x^+_{\alpha i}-\epsilon)}{Q^+(x^+_{\alpha i}+\epsilon)}\frac{Q^-(x^+_{\alpha i}-\epsilon)}{Q^-(x^+_{\alpha i}+\epsilon)}, \\
-1
& ={\fq}\frac{Q^+(x^-_{\beta i}-\epsilon)}{Q^+(x^-_{\beta i}+\epsilon)}\frac{Q^-(x^-_{\beta i}-\epsilon)}{Q^-(x^-_{\beta i}+\epsilon)},
\end{aligned}
\end{equation}
with the definition of $Q^\pm$-function given as:
\begin{align}\label{Q pm}
    Q^+(x)=\prod_{\alpha=1}^{n_+}\prod_{i=1}^\infty(x-x^{+}_{\alpha i});\quad Q^-(x)=\prod_{\beta=1}^{n_-}\prod_{i=1}^\infty(x-x^{-}_{\beta i}).
\end{align}

In order to write down $T$-function associated with this system, we define  
\begin{equation}\label{y+ and y-}
y^+(x)=\frac{Q^+(x)}{Q^+(x-\epsilon)};\quad y^-(x)=\frac{Q^-(x-\eps)}{Q^-(x)}.
\end{equation}
Both $y^+$ (resp. $y^-$) becomes degree $n_+$ (resp. $n_-$) polynomial when $x_{\alpha i}^{\pm}=\mathring{x}_{\alpha i}^{\pm}$. 
In particular in such a limit, one may identify that $y^\pm(x)$ form the characteristic polynomials of block diagonal elements of adjoint scalar $\Phi=\Phi_+\oplus\Phi_-$, i. e. 
\begin{subequations}
\begin{align}
    y^+(x)&=\prod_{\alpha=1}^{n_+}\prod_{i=1}^\infty\frac{x-a_\alpha^+-(i-1)\eps}{x-a_\alpha^+-(i-1)\eps-\eps}=\prod_{\alpha=1}^{n_+}(x-a_\alpha^+)=\det(x-\Phi_+), \\
    y^-(x)&=\prod_{\alpha=1}^{n_-}\prod_{i=1}^\infty\frac{x-a_\alpha^-+i\eps-\eps}{x-a_\alpha^-+i\eps}=\prod_{\alpha=1}^{n_-}(x-a_\alpha^-)=\det(x-\Phi_-).
\end{align}
\end{subequations}
The saddle point equations can be expressed in terms of $y^\pm(x)$ function:
\begin{subequations}\label{Saddle +-}
\begin{align}
    0&=1+\fq\frac{y^-(x^+_{\alpha i})y^-(x^+_{\alpha i}+\epsilon)}{y^+(x^+_{\alpha i})y^+(x^+_{\alpha i}+\epsilon)} \label{Saddle +},\\
    0&=1+{\fq}\frac{y^-(x^-_{\beta j})y^-(x^-_{\beta j}+\epsilon)}{y^+(x^-_{\beta j})y^+(x^-_{\beta j}+\epsilon)}\label{Saddle -}.
\end{align}
\end{subequations}
To find the resultant Bethe ansatz equation, we consider the twisted superpotential arising from the NS-limit 
\begin{align}
    \cW=\lim_{\epsilon_2\to0} \eps_2 \log[\cZ^\text{inst}_{U(n_+|n_-)}]=\cW_\text{classical}+\cW_\text{1-loop}+\cW_\text{inst}.
\end{align}
The classical twisted superpotential is given by
\begin{align}
    \cW_\text{classical}=-\log\fq\sum_{\alpha=1}^{n_+}\frac{(a_\alpha^+)^2}{2\eps}-\log\fq^{-1}\sum_{\beta=1}^{n_-}\frac{(a_\beta^-)^2}{2\epsilon},
\end{align}
and the one loop twisted superpotential is of the form
\begin{align}
    \cW_\text{1-loop}=\sum_{\sigma,\sigma'=\pm}\frac{1}{2}\sum_{(\gamma i)\neq(\gamma' i')}f(\mathring{x}^{\sigma}_{\gamma i}-\mathring{x}^{\sigma'}_{\gamma' i'}-\epsilon)-f(\mathring{x}^{\sigma}_{\gamma i}-\mathring{x}^{\sigma'}_{\gamma' i'}+\epsilon),
\end{align}
where $\mathring{x}^{\sigma}_{\gamma i}\in\{\mathring{x}^{+}_{\alpha i},\mathring{x}^{-}_{\beta i}\}$. Last but not least, there is the non-perturbative twisted superpotential
\begin{align}
    \cW_\text{inst}=\cH_\text{inst}=\sum_{\sigma,\sigma'=\pm}\mathcal{U}^{\sigma\sigma'}([x^\sigma],[x^{\sigma'}])-\mathcal{U}^{\sigma\sigma'}([\mathring{x}^{\sigma}],[\mathring{x}^{\sigma'}])
\end{align}
where $\cU^{\sigma\sigma'}$ is defined in \eqref{cU}. The twisted superpotential satisfies the equation of motion, which is the twisted F-term condition for SUSY vacua:
\begin{equation}
\frac{1}{2\pi i}\frac{\partial \cW(u)}{\partial a^\sigma_\gamma}=n^\sigma_\gamma;\quad n_\gamma^\sigma\in\mathbb{Z}.
\end{equation}
The explicit minimization calculation shows that
\begin{subequations}
\begin{align}
    &-\frac{a^+_\alpha}{\eps{}}\log\fq+\sum_{\alpha'\neq\alpha}\log\frac{\Gamma\left(\frac{a_\alpha^+-a_{\alpha'}^+}{\epsilon}\right)}{\Gamma\left(-\frac{a_\alpha^+-a_{\alpha'}^+}{\epsilon}\right)}+\sum_\beta\log \frac{\Gamma\left(-\frac{a_\alpha^+-a_{\beta}^-}{\epsilon}\right)}{\Gamma\left(\frac{a_\alpha^+-a_{\beta}^-}{\epsilon}\right)}=2\pi in_\alpha^+ ;\\
    &\frac{a^-_\beta}{\epsilon}\log\fq+\sum_\alpha\log\frac{\Gamma\left(-\frac{a_\beta^--a_{\alpha}^+}{\epsilon}\right)}{\Gamma\left(\frac{a_\beta^--a_{\alpha}^+}{\epsilon}\right)}+\sum_{\beta'\neq\beta}\log\frac{\Gamma\left(\frac{a_\beta^--a_{\beta'}^-}{\epsilon}\right)}{\Gamma\left(-\frac{a_\beta^--a_{\beta'}^-}{\epsilon}\right)} =2\pi in_\beta^-.
\end{align}
\end{subequations}
Exponentiating both sides of the above equations yields
\begin{subequations}\label{BAE super toda}
\begin{align}
1&=
\fq^{-\frac{a_\alpha^+}{2\epsilon}}\prod_{\alpha'\neq\alpha}\frac{\Gamma\left(\frac{a_\alpha^+-a_{\alpha'}^+}{\epsilon}\right)}{\Gamma\left(-\frac{a_\alpha^+-a_{\alpha'}^+}{\epsilon}\right)} \prod_\beta\frac{\Gamma\left(-\frac{a_\alpha^+-a_{\beta}^-}{\epsilon}\right)}{\Gamma\left(\frac{a_\alpha^+-a_{\beta}^-}{\epsilon}\right)}; \\ 
1&=
{\fq^{\frac{a_\beta^-}{2\epsilon}}}\prod_{\beta'\neq\beta}\frac{\Gamma\left(\frac{a_\beta^--a_{\beta'}^-}{\epsilon}\right)}{\Gamma\left(-\frac{a_\beta^--a_{\beta'}^-}{\epsilon}\right)}  \prod_\alpha\frac{\Gamma\left(-\frac{a_\beta^--a_{\alpha}^+}{\epsilon}\right)}{\Gamma\left(+\frac{a_\beta^--a_{\alpha}^+}{\epsilon}\right)}.
\end{align}
\end{subequations}
Here we propose that \eqref{BAE super toda} are the resultant Bethe ansatz equations of super Toda lattice.

In principle, the equations \eqref{Saddle +-} are enough for us to define our polynomial $T$-function. However, we would like to pause for a moment to introduce an alternative and more elegant way to obtain \eqref{Saddle +} and \eqref{Saddle -}. The reader will see shortly the new method greatly simplifies the calculation.
%\paragraph{}

In order to obtain \eqref{Saddle +} and \eqref{Saddle -}, saddle point approximation is applied to full instanton partition function under NS-limit, such that one only considers the instanton configuration that dominates the sum of all partition in \eqref{Z-nm pure}. The resultant configuration is called \textit{saddle-point configuration}. 
This means the partition function $\mathcal{Z}_{U(n_+|n_-)}^\text{inst}$ can be approximated by
\begin{align}
    \mathcal{Z}_{U(n_+|n_-)}^\text{inst}\approx\mathbb{E}\left[-\frac{SS^*}{P_{12}^*}\right][\vec{\lambda}^+_*,\vec{\lambda}^-_*]+\cdots.
\end{align}
under NS-limit, $\{\vec{\lambda}^\pm_\ast\}$ are the saddle point configurations. To obtain the saddle point equation, we may consider a small perturbation on the limit shape Young diagram, i.~e. adding/subtracting one instanton, and require the full instanton partition function to remain invariant. Suppose an additional box is added to $\vec{\lambda}_\ast^+$ with the character associated to it denoted as $e^x$, the invariance of the full partition function requires:
\begin{align}\label{qq-like}
    0&=\mathbb{E}\left[-\frac{SS^*}{P_{12}^*}\right]-\fq_+\mathbb{E}\left[-\frac{(S+P_{12}e^x)(S+P_{12}e^x)^*}{P_{12}^*}\right] \nonumber\\
    &=\mathbb{E}\left[-\frac{SS^*}{P_{12}^*}\right]\left[1-\fq_+\mathbb{E}\left[q_{12}e^x(S-P_{12}e^x)^*+Se^{-x}\right]\right].
\end{align}

Let us define $Y$-function, which is a generating function of the UV chiral ring operators,
\begin{align}\label{Y-function}
    Y(x)
    =&\mathbb{E}\left[-e^xS^*\right]=\mathbb{E}\left[-e^x(S_+-S_-)^*\right] \nonumber \\
    =&\left[\prod_\alpha\prod_{i=1}^\infty\frac{x-x_{\alpha i}^+}{x-x_{\alpha i}^+-\epsilon}\right]\times\left[\prod_{\beta}\prod_{i=1}^\infty\frac{x-x_{\beta i}^-}{x-x_{\beta i}^--\epsilon}\right]
\end{align}
with the universal bundle $S_\pm$ in \eqref{S+-}. One may immediately recognize $Y(x)$ is related to $y^\pm$  defined in \eqref{y+ and y-} by
\begin{align}
    Y(x)=\frac{y^+(x)}{y^-(x)}=\frac{\det(x-\Phi_+)}{\det(x-\Phi_-)}=\operatorname{sdet}(x-\Phi)\Big|_\text{UV}.
\end{align}
with $\Phi$ being adjoint scalar in vector multiplet in \eqref{Lagrangian}.
We have recovered saddle point equation for $x=x_{\alpha i}^+$ in \eqref{Saddle +} as
\begin{align}\label{saddle +}
    0=1+\frac{\fq_+}{Y(x^+_{\alpha i})Y(x^+_{\alpha i}+\eps)}.
\end{align}
Similar procedures can be applied to adding one additional box in Young diagram $\vec{\lambda}_\ast^-$.
\begin{align}
    0&=\mathbb{E}\left[-\frac{SS^*}{P_{12}^*}\right]-\fq_-\mathbb{E}\left[-\frac{(S-P_{12}e^x)(S-P_{12}e^x)^*}{P_{12}^*}\right] \nonumber\\
    &=\mathbb{E}\left[-\frac{SS^*}{P_{12}^*}\right]\left[1-\fq_-\mathbb{E}\left[-q_{12}e^x(S-P_{12}^*e^x)^*-Se^{-x}\right]\right] 
\end{align}
and we have recovered \eqref{Saddle -} as
\begin{align}\label{saddle -}
    0=1+\fq_-Y(x_{\beta j}^-)Y(x_{\beta j}^-+\epsilon).
\end{align}
As we have shown, \eqref{saddle +} and \eqref{saddle -} successfully reproduced \eqref{Saddle +} and \eqref{Saddle -}.
Using the fact that the coupling constants are subjects to $\fq_+=\fq_-^{-1}=\fq$, we may rewrite \eqref{saddle +} and \eqref{saddle -} into a single equation
\begin{align}
    0=1+\frac{\fq}{Y(x_{\gamma i}^\sigma)Y(x_{\gamma i}^\sigma+\eps)};\qquad x_{\gamma i}^\sigma=\{x^+_{\alpha i},x_{\beta j}^-\}. 
\end{align}

We can now define the $T$-function as
\begin{align}\label{supa-toda}
    T(x)=Y(x+\eps)+\frac{\fq}{Y(x)}=\operatorname{sdet}(x-\Phi)\Big|_\text{IR}.
\end{align}
which is identified as the super characteristic function of the $U(n_+|n_-)$ adjoint scalar $\Phi$ in IR. In the absence of $\Omega$-deformation, by denoting $z=Y(x)$, we have also successfully recovered the SW-curve of $U(n_+|n_-)$ gauge group \cite{Dijkgraaf:2016lym}:
\begin{align}
    \operatorname{sdet}(x-\Phi)=z+\frac{\fq}{z}.
\end{align}

We need to address that $T$-function defined in \eqref{supa-toda} is subjected to the restriction $0\leq|\fq|<1$ such that the gauge coupling is real (See earlier discussion around \eqref{Lagrangian}).
The open strings with both ends on positive branes therefore have kinetic energy bounded from below and the positive branes can be regarded as the \textit{physical} ones. Conversely for $|\fq|>1$, \eqref{Lagrangian} shows that the negative branes now become physical. 
This also reflects on our definition of $T$-function, as can be seen when $|\fq|\to\infty$, the asymptotic limit of $T(x)$ becomes
\begin{align}
    T(x)\approx\fq\frac{\prod_{\beta=1}^{n_-}(x-a_\beta^-)}{\prod_{\alpha=1}^{n_+}(x-a_\alpha^+)}=\frac{\det(x-\Phi_-)}{\det(x-\Phi_+)} = \frac{1}{\operatorname{sdet}(x - \Phi)},
\end{align}
where only the negative section survives.
In the rest of our paper, we will always assume $0\leq|\fq|<1$.

\subsection{Conserved Hamiltonians}
%\paragraph{}

One important aspect of the correspondence between four dimensional $\cN=2$ supersymmetric gauge theory and one-dimensional integrable model is that the $T$-function obtained from supersymmetric gauge theory can be identified as the characteristic polynomial of the corresponding integrable model. In the case of ordinary gauge group of rank $n$, the $T$-function is a degree $n$ polynomial of the form:
\begin{align}
    T(x)=\det(x-\Phi)=x^n+h_1x^{n-1}+h_2x^{n-2}+\cdots+h_{n-1}x+h_n.
\end{align}
The coefficients $\{h_i\}_{i=1}^n$ in this polynomial are conserved Hamiltonians of the corresponding $n$ particles integrable system, see \cite{Chen:2019vvt, Nikita-Shatashvili} for examples. In the case of theory with super gauge group $U(n_+|n_-)$, its $T$-function is defined as the superdeterminant of adjoint scalar field. Replacing the determinant with the superdeterminant indicates $T$-function can be written in the form of:
\begin{align}
    T(x)=\operatorname{sdet}(x-\Phi)=\frac{T_+(x)}{T_-(x)},
\end{align}
where $T_+(x)$ (resp. $T_-(x)$) is degree $n_+$ (resp. $n_-$) polynomials. 
\begin{subequations}
\begin{align}
    T_+(x)&=x^{n_+}+c_1^{(+)}x^{n_+-1}+\cdots+c_{n_+}^{(+)}, \\
    T_-(x)&=x^{n_-}+c_1^{(-)}x^{n_--1}+\cdots+c_{n_-}^{(-)}.
\end{align}
\end{subequations}
% {\color{red} Polish the context}
Our goal is to identify the coefficients $\{c_i^{(+)},c_j^{(-)}\}$ which can be identified with the conserved Hamiltonian of the corresponding integrable system. A co-dimension two full surface defect is introduced through $\bbZ_{n_++n_-} = \bbZ/(n_++n_-)\bbZ$ orbifolding \cite{Chen:2019vvt,Nekrasov:2017gzb}. Here $\bbZ_{n_++n_-}$ orbifolding acts on the coordinates of $\bR^4=\bbC_1\times\bbC_2$ by $(\mathbf{z}_1,\mathbf{z}_2)\to(\mathbf{z}_1,\zeta\mathbf{z}_2)$ where $\zeta^{n_++n_-}=1$. 

As the orbifolding procedure commutes with the NS-limit, we will temporarily restore $\epsilon_2$, and take NS limit once orbifolding is performed. We will see shortly that restoring $\eps_2$ helps us to assign the Young diagram boxes to the representations $R_\omega$ of $\mathbb{Z}_{n_++n_-}$. 
In the case of supergroup $U(n_+|n_-)$, we perform $\mathbb{Z}_{n_++n_-}$ orbifolding action into the system to introduce the full-type surface defect \cite{Kanno:2011fw}. Such an orbifolding action is characterized by a coloring function on the indices of moduli parameters 
$c: \{\alpha\}_{\alpha=1}^{n_+} \cup \{\beta\}_{\beta=1}^{n_-} \to \mathbb{Z}_{n_+ + n_-}$, which assigns each color $\alpha$ and $\beta$ to a representation $R_\omega$ of $\mathbb{Z}_{n_++n_-}$, $\omega=0,1,\dots,n_++n_--1$ \cite{Chen:2019vvt}.
In the simplest case, $c$ is defined as
\begin{equation}\label{coloring}
	c(\alpha)=\alpha-1;\quad c(\beta)=n_++\beta-1. 
\end{equation}
For this coloring function, we will denote
$[\alpha]=\{0,\dots,n_+-1\}$ and $[\beta]=\{n_+,\dots,n_++n_--1\}$ such that $[\alpha]\cup[\beta]=\{0,\dots,n_++n_--1\}$, which is the range of the index $\omega$. 
The orbifolding also splits the instanton counting parameter $\mathfrak{q}$ into $n_++n_-$ instanton counting parameters $(\fq_\omega)_{\omega=0}^{n_++n_--1}$ for each representation $R_\omega$ of $\mathbb{Z}_{n_++n_-}$. 
$(\fq_\omega)_{\omega=0}^{n_++n_--1}$ is a collection of $n_++n_-$ non-vanishing complex numbers subjected to
\begin{align}\label{split q}
    \mathfrak{q}=\prod_{\omega=0}^{n_++n_--1}\mathfrak{q}_\omega^{p(\omega)};\quad
\end{align}
with the parity function defined by
\begin{align}\label{parity}
    p(\omega)=
    \begin{cases}
    1, & \omega\in[\alpha] \\
    -1, & \omega\in[\beta]
    \end{cases}.
\end{align}

Let us now define the coordinate system 
\begin{equation}\label{Each q_w}
	\mathfrak{q}_\omega^{p(\omega)}= e^{\mathfrak{x}_{\omega}-\mathfrak{x}_{\omega-1}}, \quad\fq_{0}=\fq e^{\mathfrak{x}_0-\mathfrak{x}_{n_++n_--1}}
\end{equation}
with
\begin{equation}\label{mf x}
	\mathfrak{x}_\omega=
	\begin{cases}
		{\rx^+_\alpha}, & c^{-1}(\omega+1)=\alpha\in\{\alpha=1,\dots,n_+\}, \\
		{\rx^-_\beta}, & c^{-1}(\omega+1)=\beta\in\{\beta=1,\dots,n_-\}.
	\end{cases}
\end{equation}
The orbifolding also splits $T$-function defined in \eqref{supa-toda} into
\begin{align}
    T(x)=\prod_{\omega=0}^{n_++n_--1}T_\omega(x)
\end{align}
where each $T_\omega(x)$ is of degree one with sub-leading powers \eqref{T large x}
\begin{align}\label{def: c_w}
    T_{\omega}(x)=Y_{\omega+1}(x+\epsilon_+)+\fq_\omega\frac{1}{Y_{\omega+1}(x+\epsilon_2)}=
    \begin{cases}
    x+\epsilon+c_{1,\omega}+\frac{c_{2,\omega}}{x}+\cdots & ,\omega\in[\alpha] \\
    \fq_\omega\left[x+c_{1,\omega}+\frac{c_{2,\omega}}{x}+\cdots\right] & ,\omega\in[\beta]
    \end{cases}.
\end{align}
Notice that we modify $T(x)$ by shifting $Y(x)\to Y(x+\epsilon_2)$. This shift is necessary when the orbifolding is introduced in order to generate the correct coefficients $\{c_{1,\omega},c_{2,\omega},\dots\}$.
Under the orbifolding, $Y(x)$ defined in \eqref{Y-function} is also split into:
\begin{align}
    Y_\omega(x)
    =&\prod_\alpha({x-a^+_\alpha})^{\delta(c(\alpha)-\omega)} \prod_{(\alpha,(i,j))\in K_\omega^+}\left[\frac{x-a^+_\alpha-i\epsilon}{x-a^+_\alpha-(i-1)\epsilon}\right]\prod_{(\alpha,(i,j))\in K_{\omega+1}^+}\left[\frac{x-a^+_\alpha-(i-1)\epsilon}{x-a^+_\alpha-i\epsilon}\right], \nonumber \\
    &\times\prod_{\beta}\frac{1}{(x-a^-_\beta)^{\delta(c(\beta)-\omega)}}\prod_{(\beta,(i,j))\in K_{\omega}^-}\left[\frac{x-a^-_\beta+i\epsilon}{x-a^-_\beta+(i-1)\epsilon}\right]\prod_{(\beta,(i,j))\in K_{\omega+1}^-}\left[\frac{x-a^-_\beta+(i-1)\epsilon}{x-a^-_\beta+i\epsilon}\right]
\end{align}
with the index function:
\begin{align}
    \delta(x)=
    \begin{cases}
    1 & x=0 \\
    0 & x\neq 0
    \end{cases},
\end{align}
and
\begin{subequations}
\begin{align}
    K^+_\omega
    &=\{(\alpha,(i,j)) \mid \alpha=1,\dots,n_+;\quad(i,j)\in\lambda_*^{+,(\alpha)};\quad\alpha+j-1\equiv\omega \ \text{mod} \ n_++n_- \},\label{Kw+} \\
    K^-_\omega
    &=\{(\beta;(i,j))\mid\beta=1,\dots,n_-,\quad (i,j)\in\lambda_*^{-,(\beta)},\quad \beta-j\equiv\omega\text{ mod }n_++n_-\}.\label{Kw-}
\end{align}
\end{subequations} 
Here $K^\pm_{\omega}$ are the collections of Young diagram boxes in $\vec{\lambda}_*^\pm$ assigned to the representation $R_\omega$ of $\bbZ_{n_++n_-}$ orbifolding. 
Taking $x \to \infty$, $Y_\omega(x)$  approaches asymptotically to:
\begin{align}\label{Y_w}
    Y_{\omega}(x)=\frac{\prod_\alpha(x-a^+_\alpha)^{\delta(c(\alpha)-\omega)}}{\prod_\beta(x-a^-_\beta)^{\delta(c(\beta)-\omega)}}
    \frac{\exp\left(\frac{\epsilon_1}{x}\nu^+_{\omega-1}+\frac{\epsilon_1}{x^2}D_{\omega-1}^++\mathcal{O}(x^{-3})\right)}{\exp\left(\frac{\epsilon_1}{x}\nu^-_{\omega-1}+\frac{\epsilon_1}{x^2}D_{\omega-1}^-+\mathcal{O}(x^{-3})\right)},
\end{align}
with the following definition of notations based on collection of Young diagrams \eqref{Kw+} and \eqref{Kw-}:
\begin{subequations}\label{orbifold data}
\begin{align}
    & k_\omega^+=|K_\omega^+|;\quad \nu^+_\omega=k^+_\omega-k^+_{\omega+1};\quad \sigma^+_\omega=\frac{\epsilon}{2}k_\omega^++\sum_{(\alpha,(i,j))\in K_\omega^+}a^+_\alpha+(i-1)\epsilon;\quad D_\omega^+=-\sigma_\omega^++\sigma_{\omega+1}^+; \\
    & k_\omega^-=|K_\omega^-|;\quad \nu^-_\omega=k^-_\omega-k^-_{\omega+1};\quad \sigma^-_\omega=\frac{\epsilon}{2}k_\omega^-+\sum_{(\beta,(i,j))\in K_\omega^-}a^-_\beta-i\epsilon;\quad D_\omega^-=-\sigma_\omega^-+\sigma_{\omega+1}^-.
\end{align}
\end{subequations}
Using \eqref{Y_w}, 
we may write  $T_\omega(x)$ \eqref{def: c_w} as
\begin{align}\label{T large x}
    T_\omega(x)=
    \begin{cases}
    x+\epsilon-a^+_{c^{-1}(\omega+1)}+\epsilon\nu_{\omega}+\frac{1}{x}\left[\frac{\epsilon^2}{2}\nu_{\omega}^2-\epsilon a^+_{c^{-1}(\omega+1)}\nu_{\omega-1}+\epsilon D_{\omega}+\fq_\omega\right]+\cdots & \omega\in[\alpha] \\
    \fq_\omega \left[x+(-\epsilon\nu_\omega-a^-_{c^{-1}(\omega+1)})+\frac{1}{x}\left[\frac{\epsilon^2}{2}\nu_{\omega}^2+\epsilon (a^-_{c^{-1}(\omega+1)}-\epsilon)\nu_{\omega}-\epsilon D_{\omega}+\frac{1}{\fq_\omega}\right]+\cdots\right] & \omega\in[\beta]
    \end{cases},
\end{align}
where $\nu_\omega=\nu_\omega^+-\nu_\omega^-$, and $D_\omega=D^+_\omega-D^-_\omega$. Let us consider the following combination instead:
\begin{align}\label{T-orbifolded}
    T(x)\implies \prod_{\omega\in[\alpha]}T_\omega(x)\times\prod_{\omega\in[\beta]}\frac{1}{\fq_\omega}T_\omega(x)
\end{align}
in order to normalize the overall coefficient to be 1.
The first conserved Hamiltonian can be found by \eqref{def: c_w}
\begin{align}\label{Toda momentum}
    h_1
    &=\sum_{\omega\in[\alpha]}c_{1,\omega}-\sum_{\omega\in[\beta]}c_{1,\omega} \nonumber\\
    &=\sum_{\omega\in[\alpha]}\left[\epsilon\nu_{\omega}-a^+_{c^{-1}(\omega+1)}\right]
    +\sum_{\omega\in[\beta]}\left[\epsilon\nu_{\omega}+a^-_{c^{-1}(\omega+1)}\right] \nonumber\\
    &=\sum_{\alpha=1}^{n_+}p_\alpha^+-\sum_{\beta=0}^{n_-}p_\beta^-. 
\end{align}
The second conserved Hamiltonian is given by
\begin{align}\label{Toda-Hamiltonian}
    h_2
    &=\sum_{\omega\in[\alpha]}c_{2,\omega}-\sum_{\omega\in[\beta]}c_{2,\omega} \nonumber\\
    &=\sum_{\omega\in[\alpha]}\left[\frac{\epsilon^2}{2}\nu_{\omega}^2-\epsilon a^+_{c^{-1}(\omega+1)}\nu_\omega+\epsilon D_\omega+\fq_\omega\right]-\sum_{\omega\in[\beta]}\left[\frac{\epsilon^2}{2}\nu_\omega^2+\epsilon (a^-_{c^{-1}(\omega+1)})\nu_\omega-\epsilon D_\omega+\frac{1}{\fq_\omega}\right]
    \nonumber \\
    &=\sum_{\omega\in[\alpha]}\left[\frac{\epsilon^2}{2}\left(p^+_{c^{-1}(\omega)}\right)^2-\frac{(a^+_{c^{-1}(\omega+1)})^2}{2}+e^{\mathfrak{x}_\omega-\mathfrak{x}_{\omega-1}}\right]+\fq e^{\mathfrak{x}_0-\mathfrak{x}_{n_++n_--1}} \nonumber \\
    &\quad-\sum_{\omega\in[\beta]}\left[\frac{\epsilon^2}{2}\left(p^-_{c^{-1}(\omega)}\right)^2-\frac{(a^-_{c^{-1}(\omega+1)})^2}{2}+e^{\mathfrak{x}_\omega-\mathfrak{x}_{\omega-1}}\right].
\end{align}
We recovered conserved charges of super-Toda lattice \cite{van1994super} as advertised. Parameters in gauge/integrable correspondence can be summarized into Table~\ref{table}. 

\begin{table}
\begin{center}
\begin{tabular}{|c|c|c|}
	\hline
	& Gauge Theory & Integrable System \\ \hline\hline
	$a^\pm$ & Coulomb Moduli & Momenta \\ \hline
	$\tau$ & Complex gauge coupling & Elliptic modulus  \\ \hline
	$\epsilon$ & $\Omega$-deformation parameter & Planck constant \\ \hline
	$n_\pm$ & Gauge group rank & Number of particles \\ \hline
	$z^\pm=e^{\rx^\pm}$ & Ratio between orbifolded couplings & exponentiated coordinates  \\
	\hline
\end{tabular}
\end{center}
\caption{Parameters in gauge/integrable correspondence}
\label{table}
\end{table}

When we promote the classical Hamiltonians to the quantum mechanical operators acting on the orbifolded partition function, we may replace $\nu_\omega$ with the operator $\hat{\nu}_\omega=\frac{\partial}{\partial\mathfrak{x}_\omega}$ by recognizing that the only coordinate dependence in orbifolded instanton partition function comes from the instanton counting parameter $\prod_\omega\fq_\omega^{k_\omega}$, $k_\omega=k^+_\omega-k^-_\omega$ such that:
\begin{align}
    {\nu}_\omega\mathcal{Z}^\text{inst}_{U(n_+|n_-)}=(k_\omega-k_{\omega+1})\mathcal{Z}^\text{inst}_{U(n_+|n_-)}=\frac{\partial}{\partial\mathfrak{x}_\omega}\mathcal{Z}^\text{inst}_{U(n_+|n_-)}
\end{align}

\subsection{Super Lax operator}
%\paragraph{}
\subsubsection{Standard Arrangement}

We have established the correspondence between super Toda system and pure SYM theory with super gauge group. In particular the $T$-function with orbifolding \eqref{T-orbifolded} reproduces the desired conserved Hamiltonians \eqref{Toda-Hamiltonian}. A super integrable model is constructed by defining Lax operator $L$ on a supermatrix, see \cite{Kac:1977em,Quella:2013oda} for reviews on Lie superalgebra.
The super version of characteristic polynomial $T(x)$ is given by
\begin{align}
    T(x)=\operatorname{sdet}(x\mathbf{1}_{n_+|n_-}-L).
\end{align}
The Lax operator of super Toda lattice has been discussed in the literature, see ~\cite{van1994super} for example.
\paragraph{}
The particle interaction in Toda lattice system is restricted to only the nearest neighbors. When the system consists of more then two types of particles, the ordering of the particles becomes a crucial property of the system. The Hamiltonians obtained in previous chapter \eqref{Toda-Hamiltonian} depend on the ordering of particles, which was determined by the coloring function \eqref{coloring}: The first $n_+$ particles are in the positive sector, followed by $n_-$ in the negative sector. Let us consider the following form of sub-structure for the super-Lax matrix \cite{van1994super} which reproduces the same particle ordering:
\begin{subequations}\label{Lax std}
\begin{align}
    A_{ij}&=
    \begin{cases}
    p^+_i & \text{if } i=j; \\
    b^+_i & \text{if } j=i+1; \\
    1 & \text{if }j=i-1 \\
    0 & \text{otherwise}.
    \end{cases} \hspace{5em} %\\
    D_{ij}= 
    \begin{cases}
    p^-_i & \text{if } i=j; \\
    b^-_i & \text{if } j=i+1; \\
    1 & \text{if }j=i-1 \\
    0 & \text{otherwise},
    \end{cases} \\[.5em]
    B_{ij}&=
    \begin{cases}
    b_{n_+}^+\xi & \text{for }i=n_+, j=1 \\
    \eta & \text{for }i=1, j=n_- \\
    0 & \text{otherwise}
    \end{cases} \qquad
    C_{ij}=
    \begin{cases}
    \eta & \text{for }i=1, j=n_+ \\
    b_{n_-}^-\xi  & \text{for }i=n_-, j=1 \\
    0 & \text{otherwise}
    \end{cases} 
\end{align}
\end{subequations}
where $p_i^{\pm}$ are the canonical momenta associated with the particles in plus/minus sector, and $\xi$ and $\eta$ are the odd graded elements under superalgebra \eqref{super odd}.
The conserved Hamiltonians of the super integrable model are given by the supertrace of the super Lax matrix to integer powers:
\begin{align}
    H_i=\operatorname{str}(L^i),\quad i=1,\dots,n_++n_-.
\end{align}
In particular the first and second conserved Hamiltonians can be found easily:
\begin{subequations}
\begin{align}
    H_1&=\operatorname{str}(L)=\operatorname{tr}(A)-\operatorname{tr}(D)=\sum_{\alpha=1}^{n_+} p_\alpha^+-\sum_{\beta=1}^{n_-} p_\beta^-, \\
    H_2&=\operatorname{str}(L^2)=\operatorname{tr}(A^2+BC)-\operatorname{tr}(CB+D^2) \nonumber\\
    &=\sum_{i=1}^{n_+-1} (p_i^+)^2+2(b_i^+)+2b_{n_+}^+\xi\eta-\sum_{i=1}^{n_--1}(p_i^-)^2+2(b_i^-)+2b_{n_-}^-\xi\eta,
\end{align}
\end{subequations}
with $b^\pm_i$ given by
\begin{subequations}
\begin{align}
    b_\alpha^+&=e^{\rx^+_{\alpha+1}-\rx^+_\alpha},\quad b_{n_+}^+=e^{\rx^-_1-\rx_{n_+}^+}, \\ b_\beta^-&=e^{\rx^-_{\beta+1}-\rx^-_\beta},\quad b_{n_-}^-=\fq e^{\rx_1^+-\rx_{n_-}^-},
\end{align}
\end{subequations}
we recover the conserved Hamiltonians in \eqref{Toda momentum} and \eqref{Toda-Hamiltonian}, which fully establish the correspondence between the super version of gauge theory and Toda integrable system.
We may also consider the spectral curve of the Lax matrix, since we are dealing with a supermatrix, the corresponding spectral curve which remains Grassmannian even, is defined by the superdeterminant (Berezinian):
\begin{align}
    0&=\text{sdet}(x\mathbf{1}_{n_+|n_-}-L)=\det(x\textbf{1}_{n_+}-A-B(x\textbf{1}_{n_-}-D)^{-1}C)\times\det((x\textbf{1}_{n_-}-D)^{-1}).
\end{align}
To calculate spectral curve, we introduce the spectral parameter $y$ in the top right corner and bottom left corner elements in $L$ by
\begin{subequations}
\begin{align}
    L_{1,n_++n_-}=B_{1n_-}=y\xi \times\det(x\textbf{1}_{n_-}-D)^2;\quad L_{n_++n_-,1}=C_{n_-1}=\frac{1}{y}b_M^-\eta \times\det(x\textbf{1}_{n_-}-D)^2.
\end{align}
\end{subequations}
Since most of the elements in sub-matrices $B$ and $C$ are zero, the only non-vanishing terms in the combination $B(x\textbf{1}_{n_-}-D)^{-1}C$ are associated with the spectral parameter $y$: 
\begin{subequations}
\begin{align}
    B_{1n_-}(x\textbf{1}_{n_-}-D)^{-1}_{n_-1}C_{1n_+}
    &={y}\xi\eta\times{\det}(x\textbf{1}_{n_-}-D), \\
    B_{n_+1}(x\textbf{1}_{n_-}-D)^{-1}_{1n_-}C_{n_-1}
    &=\frac{\xi\eta}{y}b_1^-\cdots b_M^-\times\det(x\textbf{1}_{n_-}-D), \\
    B_{1n_-}(x\textbf{1}_{n_-}-D)^{-1}_{n_-n_-}C_{n_-1}
    & ; \qquad
    B_{n_+1}(x\textbf{1}_{n_-}-D)^{-1}_{11}C_{1n_+},
\end{align}
\end{subequations}
with all the other entries are zero. 
The superdeterminant (Berezinian) can be found by:
\begin{align}\label{Super-SC}
    0=\operatorname{sdet}(x\mathbf{1}_{n_+|n_-}-L)=-y\xi\eta-\frac{\fq}{y}\xi\eta+\frac{\det(x\textbf{1}_{n_+}-A-\delta A)}{\det(x\textbf{1}_{n_-}-D)}=-y\xi\eta-\frac{\fq}{y}\xi\eta+\frac{T_+(x)}{T_-(x)},
\end{align}
with
\begin{align}
    \delta A_{ij}=
    \begin{cases}
     B_{1n_-}(x\textbf{1}_{n_-}-D)^{-1}_{n_-n_-}C_{n_-1} & \text{for }i=1, j=1, \\
    B_{n_+1}(x\textbf{1}_{n_-}-D)^{-1}_{11}C_{1n_+} & \text{for }i=n_+, j=n_+, \\
    0 & \text{otherwise}.
    \end{cases} 
\end{align}
Here $T_\pm(x)$ are degree $n_+$ and $n_-$ polynomials respectively. This agrees with the expected SW curve for 4d pure $U(n_+|n_-)$ SYM theory obtained in~\cite{Dijkgraaf:2016lym}:
\begin{align}
    \Sigma_\text{SW} = \{ (x,y) \in \mathbb{C} \times \mathbb{C}^\times \mid H(x,y) = 0 \},
\end{align}
where
\begin{align}\label{SW super}
    H(x,y) = y + \fq \, \frac{1}{y} - \frac{T_+(x)}{T_-(x)},
\end{align}
with the canonical one-form on the curve $\lambda = x \frac{dy}{y}$.
The ratio of $T_\pm(x)$ is identified with the super analog of the characteristic polynomial of the adjoint scalar field in $\mathcal{N} = 2$ vector multiplet in the IR regime:
\begin{align}
    T(x) = \frac{T_+(x)}{T_-(x)} = \operatorname{sdet}(x - \Phi)\Big|_{\text{IR}}
    \, .
\end{align}
We may therefore identify the SW curve of $U(n_+|n_-)$ gauge theory with the spectral curve after replacing Grassmannian even entries $\xi\eta$ by $1$. 
The equation \eqref{SW super} agrees with the spectral curve of Super Toda-system \cite{van1994super}. We may justify this step by realizing that Grassmannian even entries in \eqref{Super-SC} are the results of choosing a representation for supergroup and superalgebra. One may choose a different convention of supergroup/superalgebra that does not involve Grassmannian. See Appendix \ref{Superalgebra} for details.
\paragraph{}
In this subsection, we have chosen a particular order of positive and negative particles in Lax matrix in \eqref{Lax std}, however general permutation can also be considered and we will discuss about it in the next section.

\subsubsection{General Arrangement: Dynkin Diagram with Fermionic Roots}\label{Special Dynkin}
%\paragraph{}

The fact that D-brane construction of supergroup uses both positive and negative branes allows for additional arrangements for them. Different ordering of the positive and negative branes can be represented by special Dynkin diagram \cite{Beisert:2003yb, Evans:1996bu}. The vertices in Dynkin diagram, for which label fundamental roots, can be either bosonic or fermionic. The fundamental roots represented by a bosonic vertices are called even roots, while fundamental roots represented by fermionic vertices are called odd roots.
Let us consider a set of three positive and three negative branes for definiteness, their different arrangements correspond to the different Dynkin diagrams with the odd roots labeling representation under $\mathfrak{sl}(3|3)$ superalgebra.
See also~\cite{Kimura:2020lmc} for a related argument in the topological string setup.
For instance, consider the Dynkin diagram with a single odd root at the 3rd/middle vertex~\cite{Dijkgraaf:2016lym,Beisert:2003yb,Evans:1996bu}:
\begin{align}
\dynkin[root radius=.2cm, edge length=1cm]{A}{ootoo}
\end{align}
%%%%%%%%%	
The odd vertex is labeled by a circle with a cross. This Dynkin diagram represents the $(+++---)$ arrangement of three positive and three negative branes. Fermionic vertex in Dynkin diagram means the two neighboring branes have different signs of Chern-Paton factors, i.~e. either a positive brane followed by a negative brane, or the other way around. The Lax supermatrix of this arrangement is given in \eqref{Lax std}.
We may consider another permutation of positive and negative branes as $(+-+-+-)$, the Dynkin diagram labeling such an arrangement of branes is given by:
\begin{align}\label{Dynkin max crossing}
\dynkin[root radius=.2cm, edge length=1cm]{A}{ttttt}.
\end{align}
%%%%%%%
A super Toda lattice defined on the root system of $\mathfrak{su}(3|3)$ algebra is represented by Dynkin diagrams. In particular the alternating ordering of positive and negative particles set up $(+-+-+-)$ can be represented by the same Dynkin diagram \eqref{Dynkin max crossing}.

The Lax operator of such an arrangement is different from the standard permutation in \eqref{Lax std}. To write down Lax matrix, we first choose a different convention for the supervectors in $\bbC^{3|3}$. In standard convention a supermatrix can be decomposed in $2\times2$ block form \eqref{supermatrix blockform} with its fundamental vector in the same convention $(+++---)^\text{T}$. The convention associated to Dynkin diagram \eqref{Dynkin max crossing} is given by
:
\begin{align}\label{Lax max}
    L_{(+-+-+-)}=
    \left.\begin{pmatrix}
    p_1^+ & \bar{b}_1^+\xi & 0 & 0 & 0 & \eta& \\
    \eta & p_1^- & \bar{b}_1^-\xi & 0 & 0 & 0 &\\
    0 & \eta & p_2^+ & \bar{b}_2^+\xi & 0 & 0 & \\
    0 & 0 & \eta & p_2^- & \bar{b}_2^-\xi & 0 & \\
    0 & 0 & 0 & \eta & p_3^+ & \bar{b}_3^+\xi & \\
     \bar{b}_3^-\xi & 0 & 0 & 0 & \eta & p_3^- &
    \end{pmatrix}\right|_{(+-+-+-)}
\end{align}
with $\bar{b}_i^\pm=e^{\alpha^\pm_i\cdot\vec{\rx}}$, $\{\alpha^\pm_i\}$ are the fundamental roots of given Dynkin diagram. In the case of \eqref{Dynkin max crossing}, it gives:
\begin{align}
    \bar{b}_1^+=e^{\rx_1^+-\rx_1^-}; \quad \bar{b}_1^-=e^{\rx_1^--\rx_2^+}; \quad 
    \bar{b}_2^+=e^{\rx_2^+-\rx_2^-}; \quad \bar{b}_2^-=e^{\rx_2^--\rx_3^+}; \quad \bar{b}_3^+=e^{\rx_3^+-\rx_3^-}; \quad \bar{b}_3^-=\fq e^{\rx_3^--\rx_1^+}.
\end{align}

One needs to convert Lax matrix in \eqref{Lax max} to standard convention for further calculation. To achieve this, notice that a fundamental vector $\vec{v}=(a_1^+,a_1^-,a_2^+,a_2^-,a_3^+,a_3^-)^T$ in the alternating convention can be mapped to standard convention:
\begin{align}
    \begin{pmatrix}
    a_1^+ \\
    a_2^+ \\
    a_3^+ \\
    a_1^- \\
    a_2^- \\
    a_3^-
    \end{pmatrix}
    =
    \begin{pmatrix}
    1 & 0 & 0 & 0 & 0 & 0 \\
    0 & 0 & 1 & 0 & 0 & 0 \\
    0 & 0 & 0 & 0 & 1 & 0 \\
    0 & 1 & 0 & 0 & 0 & 0 \\
    0 & 0 & 0 & 1 & 0 & 0 \\
    0 & 0 & 0 & 0 & 0 & 1
    \end{pmatrix}
    \begin{pmatrix}
    a_1^+ \\
    a_1^- \\
    a_2^+ \\
    a_2^- \\
    a_3^+ \\
    a_3^-
    \end{pmatrix}
    =U
    \begin{pmatrix}
    a_1^+ \\
    a_1^- \\
    a_2^+ \\
    a_2^- \\
    a_3^+ \\
    a_3^-
    \end{pmatrix}.
\end{align}
Lax matrices in standard and alternating convention is related by the same similarity transformation using the matrix $U$:
\begin{align}\label{Lax max std}
    L_{(+++---)}=
    UL_{(+-+-+-)}U^T=
    \left.\begin{pmatrix}
    p_1^+ & 0 & 0 & \bar{b}_1^+\xi & 0 & \eta& \\
    0 & p_2^+ & 0 & \bar{b}_1^-\xi & \bar{b}_2^+\xi & 0 &\\
    0 & 0 & p_3^+ & 0 & \bar{b}_2^-\xi & \bar{b}_3^+\xi & \\
    \eta & \eta & 0 & p_1^- & 0 & 0 & \\
    0 & \eta & \eta & 0 & p_2^- & 0 & \\
     \bar{b}_3^-\xi & 0 & \eta & 0 & 0 & p_3^- &
    \end{pmatrix}\right|_{(+++---)}.
\end{align}
Its nearest ordering potential is of the form:
\begin{align}
    V(\vec{\rx}^+,\vec{\rx}^-)=e^{\rx_1^+-\rx_1^-}-e^{\rx_1^--\rx_2^+}+e^{\rx_2^+-\rx_2^-}-e^{\rx_2^--\rx_3^+}+e^{\rx_3^+-\rx_3^-}-\fq e^{\rx_3^--\rx_1^+}.
\end{align}
Notice that it is not required to have only two Grassmannian numbers $\eta$ and $\xi$ in the Lax matrix, one may assign distinct Grassmannian number to each  entries in fermionic sub-blocks.
Comparing with \eqref{Lax std}, the Lax supermatrix in \eqref{Lax max std} has total different structure, which leads to different conserving Hamiltonians:
\begin{align}
    \operatorname{sdet}(x-L)=\bar{T}(x)=\frac{\bar{T}_+(x)}{\bar{T}_-(x)}.
\end{align}
On the other hand, the five-brane web formalism verifies that the instanton partition function itself does not depend on such an ordering \cite{Kimura:2020lmc} (which can be T-dual to D3-branes), so that the spectral curve, to be identified with the SW curve, and the classical Hamiltonians are consequently independent of the brane ordering.
This implies that the apparently different classical Hamiltonians are isomorphic to each other under a suitable linear transformation, whereas the quantum Hamiltonians may depend on the brane ordering due to the surface defect insertion.
It would be interesting to understand if the different arrangements of positive and negative branes as labeled by different Dynkin diagrams can lead to different super integrable systems.

\section{``Super'' XXX Spin Chain and supergroup QCD}\label{super-XXX}
%\paragraph{}

One of the most well-established gauge/integrablity correspondence is the one relating four dimensional $\cN=2$ SQCD and XXX Heisenberg spin chain \cite{HYC:2011,Nikita-Shatashvili}. 
Here we also would like to generalize this correspondence to relate the theory with the fundamental matters charged under the super gauge group and ``super" spin chains. Here ``super'' means the integrable system corresponds to supergroup SQCD is a system consisting of both  positive and negative $\mathfrak{sl}(2)$ magnons, in terms of their respective contributions to the energy eigenvalue. This is to distinguish from what normally referred as a super spin chain in the literature, which is a spin system with only positive magnons transforming under a superalgebra, for instance $\mathfrak{sl}(n_+|n_-)$, see \cite{Orlando:2010uu,Nekrasov:2018xsb,Zenkevich:2018fzl} for more details.
%\paragraph{}

In the presence of the fundamental and anti-fundamental representation hypermultiplets, the instanton partition function of $U(n_+|n_-)$ super gauge group is of the form:
\begin{align}\label{Z, matter}
    \cZ^\text{inst}_{U(n_+|n_-)}=\sum_{(\vec{\lambda}^+,\vec{\lambda}^-)}\fq^{|\vec{\lambda}^+|-|\vec{\lambda}^-|}\mathbb{E}\left[-\frac{SS^*}{P_{12}^*}+\frac{MS^*}{P_{12}^*}+\frac{S\widetilde{M}^*}{P_{12}^*}+\frac{NN^*}{P_{12}^*}-\frac{MN^*}{P_{12}^*}-\frac{N\widetilde{M}^*}{P_{12}^*}\right]\left[\vec{\lambda}^+,\vec{\lambda}^-\right]
\end{align}
where
\begin{subequations}
\begin{align}
    &M=M_+-M_-, \quad M_\pm=\sum_{i=1}^{n_\text{f}^\pm}e^{m^\pm_i}, \\ &\widetilde{M}=\widetilde{M}_+-\widetilde{M}_-, \quad\widetilde{M}_\pm=\sum_{i=1}^{n_\text{af}^\pm}e^{\widetilde{m}_i^\pm},
\end{align} 
\end{subequations}
denote the characters of fundamental matter and anti-fundamental matter. Here $S=S_+-S_-$ is defined in \eqref{S+-}, and take flavor symmetry to be $U(2n_+|2n_-)$, namely by $n_\text{f}^\pm + n_\text{af}^\pm = 2 n_\pm$.
We have proven in Section~\ref{super-Toda} that finding the saddle point equation for instanton partition function under NS-limit is equivalent to impose the following condition:
\begin{align}
    0&=\mathbb{E}\left[-\frac{SS^*}{P_{12}^*}+\frac{MS^*}{P_{12}^*}+\frac{S\widetilde{M}^*}{P_{12}^*}\right]-\fq\mathbb{E}\left[-\frac{(S-P_{12}e^{x_{\alpha i}^+})(S-P_{12}e^{x_{\alpha i}^+})^*}{P_{12}^*}+\frac{M(S-P_{12}e^{x_{\alpha i}^+})^*}{P_{12}^*}+\frac{(S-P_{12}e^{x_{\alpha i}^+})\widetilde{M}^*}{P_{12}^*}\right] \nonumber\\
    &=\mathbb{E}\left[-\frac{SS^*}{P_{12}^*}+\frac{MS^*}{P_{12}^*}+\frac{S\widetilde{M}^*}{P_{12}^*}\right]\left[1-\fq\mathbb{E}[q_{12}e^{x_{\alpha i}^+}(S-P_{12}e^{x_{\alpha i}^+})^*+Se^{-{x_{\alpha i}^+}}-Me^{-{x_{\alpha i}^+}}-q_{12}e^{x_{\alpha i}^+}\widetilde{M}^*]\right].
\end{align}
The saddle point equation can now be written as
\begin{align}
    1+\fq\frac{A({x_{\alpha i}^+}+\eps)D({x_{\alpha i}^+})}{Y({x_{\alpha i}^+})Y({x_{\alpha i}^+}+\eps)}=0,
\end{align}
with the following definitions:
\begin{subequations}
\begin{align}
    A(x)&=\mathbb{E}[-e^x\widetilde{M}^*]=\frac{\prod_{\alpha=1}^{n_+}(x-\widetilde{m}^+_\alpha)}{\prod_{\beta=1}^{n_-}(x-\widetilde{m}^-_\beta)}=\frac{A^+(x)}{A^-(x)}; \\
    D(x)&=\mathbb{E}[-e^x M^*]=\frac{\prod_{\alpha=1}^{n_+}(x-m_\alpha^+)}{\prod_{\beta=1}^{n_-}(x-m_\beta^-)}=\frac{D^+(x)}{D^-(x)},
\end{align}
\end{subequations}
such that $A^+(x)$ and $D^+(x)$ (resp. $A^-(x)$ and $D^-(x)$) are degree $n_+$ (resp. $n_-$) polynomials.
Similarly the other saddle point equation can be found as
\begin{align}\label{Saddle Hyper}
    1+{\fq}\frac{A(x_{\beta i}^-+\eps)D(x_{\beta i}^-)}{Y(x_{\beta i}^-)Y(x_{\beta i}^-+\eps)}=0.
\end{align}
Let us define the transfer function (or T-function) $T(x)$ as:
\begin{align}\label{T-matter}
    T(x)=Y(x+\eps)+\fq\frac{A(x+\eps)D(x)}{Y(x)}=\frac{y^+(x+\eps)}{y^-(x+\eps)}+\fq\frac{A^+(x+\eps)D^+(x)}{A^-(x+\eps)D^-(x)}\frac{y^-(x)}{y^+(x)}=\operatorname{sdet}(x\mathbf{1}_{n_+|n_-}-\Phi)\Big|_\text{IR},
\end{align}
which is identified as the super characteristic function of adjoint $\Phi$ in IR. In the absence of $\Omega$-deformation, we have successfully recovered the well-known classical Seiberg-Witten curve of SQCD by setting $z=Y(x)$,
\begin{align}
    \operatorname{sdet}(x\mathbf{1}_{n_+|n_-}-\Phi)\Big|_\text{IR}=z+\frac{\fq A(x)D(x)}{z}.
\end{align}
%%%%%%%%%%%%%%%%%%%%%%%%
The correspondence between SQCD and XXX spin chain is established by the identification of saddle point equation of gauge theory with BAE of XXX spin chain \cite{Nikita-Shatashvili,Dorey:2011pa,HYC:2011}. To achieve such a goal, we impose the condition between the masses of fundamental matter and Coulomb moduli parameter by the following quantization condition: 
\begin{equation}\label{quantization}
m^+_\alpha=a^+_\alpha+L^+_\alpha\epsilon;\quad m_\beta^-+\eps=a^-_\beta-(L^-_\beta)\epsilon,
\end{equation}
where $L^+_\alpha$ and $L^-_\beta$ are non-negative integers.
The quantization condition limits the number of rows in $\vec{\lambda}^+_*$ and $\vec{\lambda}^-_*$. Under the quantization condition, $y^\pm(x)$ can be rewritten as
\begin{equation}
y^+(x)=\frac{Q^+(x)}{Q^+(x-\epsilon)}D^+(x);\quad y^-(x)=\frac{Q^-(x-\epsilon)}{Q^-(x)}D^-(x),
\end{equation}
with finite products $Q^{\pm}(x)$ given by:
\begin{equation}
Q^+(x)=\prod_{\alpha=1}^{n_+}\prod_{i=1}^{L_\alpha^+}(x-x_{\alpha i}^+);\quad Q^-(x)=\prod_{\beta=1}^{n_-}\prod_{i=1}^{L_\beta^-}(x-x^-_{\beta i}).
\end{equation}
The transfer function $T(x)$ under quantization becomes
\begin{align}\label{T-hyper}
T(x)=\frac{D^+(x+\epsilon)}{D^-(x+\eps)}\frac{Q^+(x+\epsilon)Q^-(x+\epsilon)}{Q^+(x)Q^-(x)}+\fq\frac{A^+(x+\eps)}{A^-(x+\eps)}\frac{Q^+(x-\epsilon)Q^-(x-\epsilon)}{Q^+(x)Q^-(x)},
\end{align} 
such that $T(x)$ may have poles coming from $D^-(x+\eps)$ and $A^-(x+\eps)$. These poles can be eliminated by taking
\begin{align}
    A^-(x+\eps) D^-(x+\eps) T(x) = & D^+(x+\epsilon) A^-(x+\eps) \frac{Q^+(x+\epsilon)Q^-(x+\epsilon)}{Q^+(x)Q^-(x)} \nonumber\\
    &+\fq{A^+(x+\eps)}{D^-(x+\eps)}\frac{Q^+(x-\epsilon)Q^-(x-\epsilon)}{Q^+(x)Q^-(x)}.
\end{align}
Defining the Bethe roots $u_{\gamma i}^\sigma=x^\sigma_{\gamma i}+\frac{\eps}{2}$ for $
\sigma = \pm$, $\gamma = 1,\ldots,n_\sigma$, $i=1,\ldots,L_\gamma^\sigma$, and simplify the inhomogeneities (i.e.  the mass parameters for fundamental and anti-fundamental matter) by setting $\widetilde{m}_\alpha^+=m^+_\alpha+\epsilon= m^+$ for all $\alpha=1,\dots,n_+$ and $\widetilde{m}^-_\beta=m^-_\beta+\epsilon= m^-$ for all $\beta=1,\dots,n_-$
such that \eqref{Saddle Hyper} now becomes the BAE with inhomogeneities
\footnote{In general, it is allowed to assign generic inhomogeneities for both XXX spin chain and gauge theory. }:
\begin{align}\label{Bethe}
    -\fq\frac{(u_{\gamma i}^\sigma-{m}^++\frac{\eps}{2})^{n_+}}{(u_{\gamma i}^\sigma-m^+-\frac{\eps}{2})^{n_+}}\frac{(u_{\gamma i}^\sigma-m^--\frac{\eps}{2})^{n_-}}{(u_{\gamma i}^\sigma-{m}^-+\frac{\eps}{2})^{n_-}}
    =\prod_{\{u_{\gamma' i'}^{\sigma'}\}}\frac{u_{\gamma i}^\sigma-u_{\gamma' i'}^{\sigma'}+\eps}{u_{\gamma i}^\sigma-u_{\gamma' i'}^{\sigma'}-\eps}.
\end{align}
The spin chain system that \eqref{Bethe} represents is a coupling system of length $n_+$ positive magnon and another spin chain of length $n_-$ negative magnon \cite{Kimura:2019msw}. 
We can compare this with the Bethe ansatz equation of $\mathfrak{sl}(2)$ spin-$s$ XXX spin chain of length $L$ given by \cite{Faddeev:1996iy}:
\begin{align}
    \left(\frac{u_j-m+s\hbar}{u_j-m-s\hbar}\right)^L=\prod_{k(\neq j)}^n \frac{u_j-u_k+\hbar}{u_j-u_k-\hbar},
\end{align}
where $\{u_j\}_{j=1}^n$ are Bethe roots and $m$ is inhomogeneity. The energy of this system for arbitrary spin-$s$ has been found to be:
\begin{align}
    E=-\frac{\hbar J}{2}\sum_{j=1}^n\frac{d}{du_j}\ln\frac{u_j-m+s\hbar}{u_j-m-s\hbar}=\frac{\hbar J}{2}\sum_{j=1}^n\left(\frac{1}{u_j-m-s\hbar}-\frac{1}{u_j-m+s\hbar}\right)=\frac{J}{2}\sum_{j=1}^n\frac{2s\hbar^2}{(u_j-m)^2-s^2\hbar^2}.
\end{align}
When $n_+=0$, \eqref{Bethe} coincides with BAE of $\mathfrak{sl}(2)$ spin ``$s=-\frac{1}{2}$'' system, which may be interpreted as the ``negative magnon'' excitation. However, we should remember that the BAE always comes with the product of spin and Planck constant, we can not distinguish whether the negative magnon is indeed spin $-\frac{1}{2}$, or we have assigned a negative Planck constant. As discussed in section \ref{adhm}, the positive and negative sectors in supergroup gauge theory are charged oppositely in the $\Omega$-background \eqref{ADHM character}, 
given that $\epsilon_1 =\epsilon$ is identified as the Planck constant here, it would be interesting to show whether we can use this property of the gauge theory to assign opposite valued Planck constants to different sub-sectors in a quantum integrable system.
%\paragraph{}

The energy of the total system \eqref{Bethe} is now a combination of positive and negative magnons, which can be found by turning off the inhomogeneities $m^+=m^-=0$. For the positive magnons, we have
\begin{align}
    E_+=-\frac{\eps J}{2}\sum_{\alpha=1}^{n_+}\sum_{i=1}^{L^+_\alpha}\frac{d}{du_{\alpha i}^+}\ln\frac{u_{\alpha i}^++\frac{\eps}{2}}{u_{\alpha i}^+-\frac{\eps}{2}}=\frac{J}{2}\sum_{\alpha=1}^{n_+}\sum_{i=1}^{L^+_\alpha}\frac{\epsilon^2}{(u_{\alpha i}^+)^2-\frac{\epsilon^2}{4}},
\end{align}
while the negative magnons contribute
\begin{align}
    -E_-=-\frac{\eps J}{2}\sum_{\beta=1}^{n_-}\sum_{i=1}^{L_\beta^-}\frac{d}{du^-_{\beta i}}\ln\frac{u^-_{\beta i}-\frac{\eps}{2}}{u^-_{\beta i}+\frac{\eps}{2}}=-\frac{J}{2}\sum_{\beta=1}^{n_-}\sum_{i=1}^{L_\beta^-}\frac{\epsilon^2}{(u^-_{\beta i})^2-\frac{\epsilon^2}{4}}
\end{align}
The overall minus sign in front of the energy is how negative magnons get their name. The total energy of the system is simply combination of the two contributions
\begin{align}
    E=E_+-E_-  =\frac{J}{2}\sum_{\alpha=1}^{n_+}\sum_{i=1}^{L^+_\alpha}\frac{\epsilon^2}{(u_{\alpha i}^+)^2-\frac{\epsilon^2}{4}}-\frac{J}{2}\sum_{\beta=1}^{n_-}\sum_{i=1}^{L_\beta^-}\frac{\epsilon^2}{(u_{\beta i}^-)^2-\frac{\epsilon^2}{4}}.
\end{align}
Notice that unlike Toda lattice, surface defect is introduce by quantization condition in \eqref{quantization} instead of orbifolding.
In particular in the case of SQCD, if the surface defects are introduced via orbifolding instead of quantization condition \eqref{quantization}, the same procedure gives rise not to XXX spin chain, but Gaudin model~\cite{Nekrasov:2017gzb}, which are spectral dual to each other~\cite{Mironov:2012uh,Mironov:2012ba}.

\section{Elliptic Double Calogero-Moser System from $\mathcal{N}=2^*$ supergroup Gauge Theory}\label{edCM}
%\paragraph{}

It is well known that four dimensional $\mathcal{N}=2^*$ $U(n)$ gauge group theory  is associated with elliptic Calogero-Moser (eCM) system \cite{Chen:2019vvt, Nekrasov:2017gzb, Nikita-Shatashvili} whose Hamiltonian is defined on the root system $R$ of given algebra~\cite{olshanetsky1983quantum}: 
\begin{align}\label{eCM root}
    H=\Delta+\sum_{\alpha\in R^+}k_\alpha(k_\alpha-1){(\alpha,\alpha)}{\wp(\alpha\cdot \vec{x};\tau)}
\end{align}
where $\Delta$ is the Laplacian associated with the algebra, $R^+$ denotes the positive root system, and $\wp(x)$ is the Weierstrass $\wp$-function with the definition~\eqref{Def:p-function}. 
The super Lax formalism generalization of Calogero-Moser system was studied in \cite{sergeev2001superanalogs,HiJack}, which is called the double Calogero-Moser (dCM) system by considering the root system associated with the superalgebra. Let $\{e_i^+,e_j^-\}$ be the orthonormal basis of $\mathbb{C}^{n_+|n_-}$, such that $(e_i^\sigma,e_j^{\sigma'}) = \delta_{ij} \delta_{\sigma \sigma'}$. 
The root system $R$ of superalgebra $U(n_+|n_-)$ can be described by $R = \coprod_{\sigma,\sigma' = \pm} R_{\sigma\sigma'}$ %$R=R_{++}\amalg R_{--}\amalg R_{+-}\amalg R_{-+}$ 
with
\begin{subequations}
\begin{align}
    & R_{++}=\{e_i^+-e_j^+|i,j=1,\dots,n_+\},\quad 
     R_{--}=\{e^-_i-e^-_j|i,j=1,\dots,n_-\}, \\
    & R_{+-}=\{e_i^+-e_j^-|i=1,\dots,n_+; j=1,\dots,n_-\},\quad 
     R_{-+}=\{e^-_i-e^+_j|i=1,\dots,n_-; j=1,\dots,n_-\}.
\end{align}
\end{subequations}
If one simply extends \eqref{eCM root} to the root system of superalgebra, any interactions labeled by the root $\gamma=e^+_i-e^-_j\in R_{+-}$ for some $i=1,\dots,n_+$, $j=1,\dots,n_-$ must naturally vanish by inner product $(\gamma,\gamma)=(e_i^+,e^+_i)-(e^-_j,e^-_j)=1-1=0$. 
To resolve such a problem hence facilitate the non-trivial interactions, the author of \cite{sergeev2001superanalogs} introduced an deformation on the inner product of root system. Let $v$ and $w$ be vector in $\mathbb{C}^{n_+|n_-}$, define
\begin{align}\label{inner product}
    (v,w)_\theta=\sum_{i=1}^{n_+}v(e_i^+)w(e_i^+)-\theta\sum_{j=1}^{n_-}v(e_j^-)w(e_j^-),
\end{align}
this deformation of the inner product parameterized by $\theta$ explicitly breaks the supergroup $U(n_+|n_-)\to U(n_+)\times U(n_-)$. The Hamiltonian of edCM consists of two sets of identical particles, given by:
\begin{align}\label{Hamiltonian edCM}
    H=&-\frac{1}{2}\sum_{\alpha=1}^{n_+}\frac{\partial^2}{\partial(\rx_\alpha^+)^2}+\frac{k}{2}\sum_{\beta=1}^{n_-}\frac{\partial^2}{\partial(\rx_\beta^-)^2} \nonumber \\
    &+k(k-1)\sum_{\alpha>\alpha'}\wp(\rx_\alpha^+-\rx_{\alpha'}^+;\tau)+(1-k)\sum_{\alpha,\beta}\wp(\rx_\alpha^+-\rx_\beta^-;\tau)+(1-\frac{1}{k})\sum_{\beta>\beta'}\wp(\rx_\beta^--\rx_{\beta'}^-;\tau).
\end{align}
The Hamiltonian in \eqref{Hamiltonian edCM} is characterized by a single coupling constant $k$ and is identified with the inner product parameter $\theta=k$.
The integrability of trigonometric dCM system was proved in \cite{HiJack}, and the integrability of elliptic potential was proven in \cite{Chen:2019vvt}. 
The gauge theory associated to edCM was constructed in \cite{Nekrasov:2017gzb, Chen:2019vvt} using so-called gauge origami construction for the ordinary gauge group. 
However based on the original proposal made in \cite{sergeev2001superanalogs} for the supersymmetric analogue of edCM, one should also be able to find the similar correspondence with supergroup gauge theory in four dimensions. In particular, we would like to reproduce conserved Hamiltonians of edCM from the q-function of $\cN=2^*$ $U(n_+|n_-)$ super gauge group theory in this section.

\subsection{Saddle point equation from instanton partition function}
%\paragraph{}

The instanton partition function of four dimensional $\cN=2^*$ $U(n_+|n_-)$ gauge theory can be written as:
\begin{align}\label{Z inst adj}
    \cZ^{\cN=2^*}_{U(n_+|n_-),\text{inst}}
    &=\sum_{(\vec{\lambda}^+,\vec{\lambda}^-)}\fq^{|\vec{\lambda}^+|-|\vec{\lambda}^-|}\mathbb{E}\left[-(1-e^m)\frac{SS^*-NN^*}{P_{12}^*}\right][\vec{\lambda}^+,\vec{\lambda}^-] \nonumber\\
    &=\sum_{(\vec{\lambda}^+,\vec{\lambda}^-)}\fq^{|\vec{\lambda}^+|-|\vec{\lambda}^-|}\cZ_\text{adj}^{++}\cZ_\text{adj}^{+-}\cZ_\text{adj}^{-+}\cZ_\text{adj}^{--},
\end{align}
where $m$ now denotes the mass of adjoint hypermultiplet. The universal bundle $S=S_+-S_-$ is defined in \eqref{S+} and \eqref{S-} and $N=N_+-N_-$, $N_\pm=\sum_{\alpha=1}^{n_\pm} e^{a^\pm_\alpha}$. Following the similar calculation to pure SYM case, one finds:
\begin{align}
    \mathcal{Z}^{\sigma\sigma'}_\text{adj}[\vec{\lambda}^+,\vec{\lambda}^-]
    &=\left[\left.\prod_{\alpha=1}^{n_\sigma}\prod_{\beta=1}^{n_{\sigma'}}\prod_{i=1}^\infty\prod_{j=1}^\infty\right|_{(\alpha i)\neq(\beta j)} \frac{\Gamma(\epsilon_2^{-1}(x^\sigma_{\alpha i}-x^{\sigma'}_{\beta j}-\epsilon_1))}{\Gamma(\epsilon_2^{-1}(x^\sigma_{\alpha i}-x^{\sigma'}_{\beta j}))}\cdot\frac{\Gamma(\epsilon_2^{-1}(x^\sigma_{\alpha i}-x^{\sigma'}_{\beta j}-m))}{\Gamma(\epsilon_2^{-1}(x^\sigma_{\alpha i}-x^{\sigma'}_{\beta j}-\epsilon_1-m))}\right]\nonumber, \\
    & \times\left[\left.\prod_{\alpha=1}^{n_\sigma}\prod_{\beta=1}^{n_{\sigma'}}\prod_{i=1}^\infty\prod_{j=1}^\infty\right|_{(\alpha i)\neq(\beta j)}
    \frac{\Gamma(\epsilon_2^{-1}(\mathring{x}_{\alpha i}^{\sigma}-\mathring{x}_{\beta j}^{\sigma'}))}{\Gamma(\epsilon_2^{-1}(\mathring{x}_{\alpha i}^{\sigma}-\mathring{x}_{\beta j}^{\sigma}-\epsilon_1))}\cdot\frac{\Gamma(\epsilon_2^{-1}(\mathring{x}_{\alpha i}^{\sigma}-\mathring{x}_{\beta j}^{\sigma}-\epsilon_1-m))}{\Gamma(\epsilon_2^{-1}(\mathring{x}_{\alpha i}^{\sigma}-\mathring{x}_{\beta j}^{\sigma'}-m))}\right],
\end{align}
where $\sigma, \sigma' = \pm$.
As before, we consider NS-limit $\epsilon_2\to0$ with $\epsilon_1:=\epsilon$ fixed. By using Stirling approximation of $\Gamma$-function, the instanton partition function can be written as
\begin{align}
    \cZ_\text{adj}^{\sigma\sigma'}[\vec{\lambda}^+,\vec{\lambda}^-]\approx\exp\left[\frac{1}{2\epsilon_2}\sum_{(\gamma i)\neq(\gamma' i')}\right.
    & 	f(x^\sigma_{\gamma i}-x^{\sigma'}_{\gamma' i'}-\epsilon)-f(x^\sigma_{\gamma i}-x^{\sigma'}_{\gamma' i'}+\epsilon) \nonumber \\
    & +f(x^\sigma_{\gamma i}-x^{\sigma'}_{\gamma' i'}-m)-f(x^\sigma_{\gamma i}-x^{\sigma'}_{\gamma' i'}+m)\nonumber \\
    &-f(x^{\sigma}_{\gamma i}-x^{\sigma'}_{\gamma' i'}-m-\epsilon)+f(x^\sigma_{\gamma i}-x^{\sigma'}_{\gamma' i'}+m+\epsilon) \nonumber\\
    -\frac{1}{2\epsilon_2}\sum_{(\gamma i)\neq(\gamma' i')}
    & 	f(\mathring{x}^{\sigma}_{\gamma i}-\mathring{x}^{\sigma'}_{\gamma' i'}-\epsilon)-f(\mathring{x}^{\sigma}_{\gamma i}-\mathring{x}^{\sigma'}_{\gamma' i'}+\epsilon) \nonumber \\
    & +f(\mathring{x}^{\sigma}_{\gamma i}-\mathring{x}^{\sigma'}_{\gamma' i'}-m)-f(\mathring{x}^{\sigma}_{\gamma i}-\mathring{x}^{\sigma'}_{\gamma' i'}+m)\nonumber \\
    &\left.-f(\mathring{x}^{\sigma}_{\gamma i}-\mathring{x}^{\sigma'}_{\gamma' i'}-m-\epsilon)+f(\mathring{x}^{\sigma}_{\gamma i}-\mathring{x}^{\sigma'}_{\gamma' i'}+m+\epsilon)\right].
\end{align}
where $f(x)=x(\log x-1)$. The discrete sum in the instanton partition function above becomes an integration over set of infinite integration variables $\{x^+_{\alpha i},x^-_{\beta i}\}$ defined in \eqref{x data}:
\begin{equation}\label{Z-int}
\begin{aligned}
\cZ^{\cN=2^*}_{U(n_+|n_-)}
& =\int\prod_{\alpha i}dx^+_{\alpha i}\prod_{\beta i}dx^-_{\beta i}\exp\left[\frac{1}{\epsilon_2}\mathcal{H}_\text{inst}([x^+],[x^-])\right] \\
&=\int\prod_{\alpha i}dx^+_{\alpha i}\prod_{\beta i}dx^-_{\beta i}\exp\left[\frac{1}{\epsilon_2}\sum_{\sigma,\sigma'=\pm}\mathcal{H}_\text{inst}^{\sigma\sigma'}([x^\sigma],[x^{\sigma'}])+\frac{1}{\epsilon_2}\sum_{\sigma=\pm}\mathcal{H}_\text{inst}^\sigma([x^\sigma])\right]
\end{aligned}
\end{equation}
where
\begin{equation}
\begin{aligned}
& \mathcal{H}_\text{inst}^{\sigma\sigma'}([x^\sigma],[x^{\sigma'}])=\mathcal{U}^{\sigma\sigma'}([x^\sigma],[x^{\sigma'}])-\mathcal{U}^{\sigma\sigma'}([\mathring{x}^{\sigma}],[\mathring{x}^{\sigma'}]); \\
&\mathcal{H}_\text{inst}^\sigma([x^{\sigma}])=\mathcal{V}^\sigma([x^{\sigma}])-\mathcal{V}^\sigma([\mathring{x}^{\sigma}]);\qquad \mathcal{V}^\sigma([x^\sigma])=\sigma\log\fq\sum_{(\alpha i)}x^{\sigma}_{\alpha i}.
\end{aligned}
\end{equation}
The function $\cU^{\sigma\sigma'}$ is of the following form
\begin{equation}\label{cU adj}
\begin{aligned}
\mathcal{U}^{\sigma\sigma'}([x^\sigma],[x^{\sigma'}])=\frac{1}{2}\sum_{(\gamma i)\neq(\gamma' i')}
& 	f(x^\sigma_{\gamma i}-x^{\sigma'}_{\gamma' i'}-\epsilon)-f(x^\sigma_{\gamma i}-x^{\sigma'}_{\gamma' i'}+\epsilon) \\
& +f(x^\sigma_{\gamma i}-x^{\sigma'}_{\gamma' i'}-m)-f(x^\sigma_{\gamma i}-x^{\sigma'}_{\gamma' i'}+m) \\
& -f(x^\sigma_{\gamma i}-x^{\sigma'}_{\gamma' i'}-m-\epsilon)+f(x^\sigma_{\gamma i}-x^{\sigma'}_{\gamma' i'}+m+\epsilon). \\
\end{aligned}
\end{equation}
%%%%%%%%%%%%%%%%%%%
To calculate the integration in \eqref{Z-int}, we introduce the instanton density $\rho(x)$ which equals to unity along $\mathfrak{J}=\mathfrak{J}^+\cup\mathfrak{J}^-$, $\mathfrak{J}^+=\bigcup_{\alpha i}[\mathring{x}^{+}_{\alpha i},x^+_{\alpha i}]$, $\mathfrak{J}^-=\bigcup_{\beta i}[\mathring{x}^{-}_{\beta i},x^-_{\beta i}]$, and vanishes otherwise. The contributions in \eqref{Z-int} can be expressed as
\begin{equation}
\mathcal{H}^{\sigma\sigma'}=-\frac{1}{2}\int_{\mathfrak{J}^\sigma}dx\int_{\mathfrak{J}^{\sigma'}}dy\,{G}(x-y)+\int_{\mathfrak{J}^\sigma}dx \log({R}^{\sigma'})
\end{equation}
where
\begin{equation}
{G}(x)=\frac{d}{dx}\log\frac{(x+m+\epsilon)(x-m)(x-\epsilon)}{(x-m-\epsilon)(x+m)(x+\epsilon)};\qquad{R}^{\sigma'}(x)=\frac{P^{\sigma'}(x-m)P^{\sigma'}(x+m+\epsilon)}{P^{\sigma'}(x)P^{\sigma'}(x+\epsilon)}.
\end{equation}
%%%%%%%%%%%%%%%
The integration in $\mathcal{H}_\text{inst}$ is dominated by the saddle point configuration, which can be obtained from varying $\cH_{\rm inst}$ with respect to $x^+_{\alpha i}$ or $x^-_{\beta i}$:
\begin{equation}
\begin{aligned}
0=\frac{\delta\mathcal{H}_\text{inst}}{\delta x^+_{\alpha i}}
&=\frac{\delta\mathcal{H}^{++}}{\delta x^+_{\alpha i}}+\frac{\delta\mathcal{H}^{+-}}{\delta x^+_{\alpha i}}+\frac{\delta\mathcal{H}^{-+}}{\delta x^+_{\alpha i}}+\frac{\delta\mathcal{H}^{+}}{\delta x^+_{\alpha i}}, \\
&=-\int_{\mathfrak{J}^+} dy\,{G}(x^+_{\alpha i}-y)+\int_{\mathfrak{J}^-}dy\,{G}(x^+_{\alpha i}-y)+\log({R}^+(x^+_{\alpha i}))-\log({R}^-(x^+_{\alpha i}))+\log\fq, \\
0=\frac{\delta\mathcal{H}_\text{inst}}{\delta x^-_{\beta i}}
&=\frac{\delta\mathcal{H}^{+-}}{\delta x^-_{\beta i}}+\frac{\delta\mathcal{H}^{-+}}{\delta x^-_{\beta i}}+\frac{\delta\mathcal{H}^{--}}{\delta x^-_{\beta i}}+\frac{\delta\mathcal{H}^{-}}{\delta x^-_{\beta i}} \\
&=+\int_{\mathfrak{J}^+} dy\,{G}(x^-_{\beta i}-y)-\int_{\mathfrak{J}^-}dy\,{G}(x^-_{\beta i}-y)-\log({R}^+(x^-_{\beta i}))+\log({R}^-(x^-_{\beta i}))+\log\fq.
\end{aligned}
\end{equation}
Since ${G}(x)$ is a total derivative, we can rewrite the defining equation for the saddle point configuration into
\begin{equation}\label{BAE-super}
\begin{aligned}
-1
& =\fq\frac{Q^+(x^+_{\alpha i}+m+\epsilon)Q^+(x^+_{\alpha i}-m)Q^+(x^+_{\alpha i}-\epsilon)}{Q^+(x^+_{\alpha i}-m-\epsilon)Q^+(x^+_{\alpha i}+m)Q^+(x^+_{\alpha i}+\epsilon)}
\frac{Q^-(x^+_{\alpha i}+m+\epsilon)Q^-(x^+_{\alpha i}-m)Q^-(x^+_{\alpha i}-\epsilon)}{Q^-(x^+_{\alpha i}-m-\epsilon)Q^-(x^+_{\alpha i}+m)Q^-(x^+_{\alpha i}+\epsilon)}, \\
-1
& ={\fq}\frac{Q^+(x^-_{\beta i}+m+\epsilon)Q^+(x^-_{\beta i}-m)Q^+(x^-_{\beta i}-\epsilon)}{Q^+(x^-_{\beta i}-m-\epsilon)Q^+(x^-_{\beta i}+m)Q^+(x^-_{\beta i}+\epsilon)}
\frac{Q^-(x^-_{\beta i}+m+\epsilon)Q^-(x^-_{\beta i}-m)Q^-(x^-_{\beta i}-\epsilon)}{Q^-(x^-_{\beta i}-m-\epsilon)Q^-(x^-_{\beta i}+m)Q^-(x^-_{\beta i}+\epsilon)}.
\end{aligned}
\end{equation}
with $Q^\pm(x)$ defined in \eqref{Q pm}. 
The two saddle point equations in \eqref{BAE-super} can be further packaged into a single equation using $Y(x)$ defined in \eqref{Y-function} as:
\begin{align}\label{saddle Y adj}
    1+\fq\frac{Y(x^\sigma_{\gamma i}+m+\epsilon)Y(x^\sigma_{\gamma i}-m)}{Y(x^\sigma_{\gamma i})Y(x^\sigma_{\gamma i}+\epsilon)}=0.
\end{align}

\subsection{$\mathbb{X}$-function and Conserved Charges}
%\paragraph{}

One major difference between $\cN=2^*$ theory and the other cases we have considered earlier in this work is that the $T$-function of $\cN=2^*$ system $\bbX(x)$, is an infinite series \cite{Nekrasov:2012xe,Nekrasov:2013xda,Feigin:2017wnq,Nekrasov:2017gzb}. It is called the $qq$-character for generic $\epsilon_{1,2}$~\cite{Nekrasov:2015wsu,Bourgine:2015szm,Kimura:2015rgi} or the $q$-character in the NS limit, $\epsilon_2 \to 0$.
This is because $\cN=2^*$ theory is classified as an affine quiver theory of the type $\widehat{A}_0$, whereas the others are of $A_1$ quiver.
The function $\bbX(x)$ for $U(n_+|n_-)$ group gauge theory (with generic $\epsilon_{1,2}$ and without $\theta$-deformation) can be written as~\cite{Kimura:2019msw}:
\begin{align}\label{Xbb+}
    &\mathbb{X}(x)={Y(x+\epsilon_+)}\sum_{\{{\mu}\}}\fq^{|{\mu}|}{B}[{\mu}]\prod_{(\bi,\bj)\in\mu}\frac{Y(x+s_{\bi\bj}-m)Y(x+s_{\bi\bj}+m+\epsilon_+)}{Y(x+s_{\bi\bj})Y(x+s_{\bi\bj}+\epsilon_+)} ,
\end{align}
where $\epsilon_+=\epsilon_1+\epsilon_2$.  
Here $\mu$ is a single auxiliary partition denoted by
\begin{align}
    \mu=(\mu_1,\mu_2,\dots,\mu_{\ell(\mu)}).
\end{align}
Notice that $\mu$ has no relation to $(\vec{\lambda}^\pm)$ which label the equivariant fix points on the instanton moduli space.%
\footnote{%
One could interpret the partition $\mu$ as describing the instanton configuration on the transverse subspace in $\mathbb{C}^2 \times \mathbb{C}^2 = \mathbb{C}^4$, called the gauge origami~\cite{Nekrasov:2015wsu,Chen:2019vvt}}
Since $\mu$ denotes only one Young diagram, we will not use a vector notation. Each box in $\mu$ is labeled by
\begin{align}
    s_{\bi\bj}=(\bi-1)m-(\bj-1)(m+\epsilon_+)
\end{align}
where $\bi=1,\dots,\ell(\mu)$ and $\bj=1,\dots,\mu_\bi$ with given $\bi$. We also define
\begin{align}\label{B[mu]}
    B[\mu]=\prod_{(\bi,\bj)\in\mu}B_{1,2}(mh_{\bi\bj}+\epsilon_+);\quad B_{1,2}(x)=1+\frac{\epsilon_1\epsilon_2}{x(x+\epsilon_+)}
\end{align}
from $\mu$.
Here $a_{\bi\bj}=\mu_\bi-\bj$ denotes the ``arm'' of associated box $(\bi,\bj)$ in Young diagram $\mu$, and $l_{\bi\bj}=\mu^\text{T}_\bj-\bi$ is the ``leg'' associated to the same box. 
We have also defined the hook length $h_{\bi\bj}=a_{\bi\bj}+l_{\bi\bj}+1$. Under NS limit $\epsilon_2\to0$ which most of our works consider, $B[\mu]=1$ for all $\mu$ by \eqref{B[mu]}.

To find the conserved Hamiltonians arising from our $\cN=2^*$ theory with super gauge group, a $\bbZ_{n_++n_-}$ orbifolding is further introduced in the same way we had done for super Toda lattice. 
We will use the same coloring function as in \eqref{coloring}. 
Under the orbifolding, $\mathbb{X}(x)$ function in \eqref{Xbb+} split into $n_++n_-$ copies:
\begin{align}
    \mathbb{X}(x)=\prod_{\omega=0}^{n_++n_--1}\mathbb{X}_{\omega}(x),
\end{align}
with each
\begin{align}
    \mathbb{X}_\omega(x)=Y_{\omega+1}(x+\epsilon_+)\sum_{\{\mu\}}\mathbb{B}_\omega^\mu (\vec{\mathfrak{z}}, \tau) \prod_{(\bi,\bj)\in\mu}\frac{Y_{\omega+1-\bj}(x+s_{\bi\bj}-m)Y_{\omega+1-\bj+1}(x+s_{\bi\bj}+m+\epsilon_+)}{Y_{\omega+1-\bj}(x+s_{\bi\bj})Y_{\omega+1-\bj+1}(x+s_{\bi\bj}+\epsilon_+)} \label{Xbb_w}.
\end{align}
Here we would like to address the $\mathbb{B}_\omega^\mu$. It is the orbifolded $\fq^{|\mu|}B[\mu]$ defined in \eqref{B[mu]}, each given by
\begin{align}
    \mathbb{B}^{\mu}_\omega=\left.\prod_{(\bi,\bj)\in\mu}\mathfrak{q}_{\omega+1-\bj}B_{1,\omega}((-1)^{p(\omega+1-\bj)}mh_{\bi\bj} )\right|_{a_{\bi\bj}=0}
\end{align}
with 
\begin{equation}
B_{1,\omega}(x)=1+(-1)^{p(\omega)}\frac{\epsilon}{x},
\end{equation}
where $p(\omega)$ is parity function defined in \eqref{parity}
\paragraph{}
One way to think about this configuration is that the orbifolding now splits the instanton partition into $n_++n_-$ copies of $U(1)$ sub-partitions. Each element in $K_\omega$ is counted by orbifolded coupling $\mathfrak{q}_\omega$ instead of the original $\mathfrak{q}$. To evaluate the summation over all possible Young diagrams, we will introduce a new representation for a Young diagram $\mu$:
\begin{equation}
    \mu=(1^{l_0}2^{l_1}\dots (N-1)^{l_{N-2}}(N)^l).
\end{equation}
Each $l_{r-1}=\sum_{J=0}^\infty l_{r-1,J}$, where $l_{r-1,J}=\left(\mu_{r+NJ}^{T}-\mu^T_{r+1+NJ}\right)$ is the difference between number of boxes of two neighboring columns, $r = 1,\ldots, N-1$ and the last one $\bl=\sum_{J=1}^\infty\mu_{NJ}^T$ counts for how many times a full combination of $\mathfrak{q}_0\cdots\mathfrak{q}_{N-1}=\mathfrak{q}$ shows up.
Defining the summation over all possible partition configurations of each $\omega$ as:
Summing over all possible Young diagrams $\{\mu\}$ gives:
\begin{equation}\label{B-form}
\begin{aligned}
    \mathbb{B}_\omega(\vec{\mathfrak{z}};\tau)
    &=\sum_{\{\mu\}}\mathbb{B}_\omega^{\mu}(\vec{\mathfrak{z}};\tau)
    =\sum_{l_0,\dots,l_{N-1},\bl\geq0}\prod_{\alpha=0}^{N-1}\frac{\left((-1)^{p(\omega)+p(\alpha)}l_\alpha+\frac{\epsilon_1}{m}\right)!}{(l_\alpha)!\left(\frac{\epsilon_1}{m}\right)!}\left(\frac{\mathfrak{z}_\omega}{\mathfrak{z}_\alpha}\right)^{l_\alpha}\mathfrak{q}^\bl \\
    &=\prod_{\omega'<\omega}\left[\frac{1}{(\frac{\mathfrak{z}_\omega}{\mathfrak{z}_{\omega'}};\fq)_\infty}\right]^{(-1)^{p(\omega)+p(\omega')}\frac{m+\eps}{m}}\prod_{\omega'\geq\omega}\left[\frac{1}{(\fq\frac{\mathfrak{z}_\omega}{\mathfrak{z}_{\omega'}};\fq)_\infty}\right]^{(-1)^{p(\omega)+p(\omega')}\frac{m+\eps}{m}},
\end{aligned}
\end{equation}
with $\mathfrak{z}_\omega=e^{\mathfrak{x}_\omega}$, $\mathfrak{x}_\omega$ was defined in \eqref{mf x}.
The $q$-Pochhammer function $(z;q)_n$ is defined in \eqref{q-Pochhammer}, the total product of $\mathbb{B}_\omega$ is given by:
\begin{equation}\label{B=QF}
    \mathbb{B}(\vec{\mathfrak{z}}, \tau)=\prod_\omega\mathbb{B}_\omega (\vec{\mathfrak{z}}, \tau)=\mathbb{Q}^{\frac{m+\epsilon}{m}}(\vec{\mathfrak{z}};\tau).
\end{equation}
The function $\mathbb{Q}$ consists of three parts:
$\mathbb{Q}=\mathbb{Q}_{++}\mathbb{Q}_{+-}^{-1}\mathbb{Q}_{--}$, where each element is given by
\begin{subequations}
\begin{align}
    \mathbb{Q}_{++}^{-1}
    &=\left[\prod_{n_+\geq\alpha>\alpha'\geq1}\frac{\theta_{11}\left(\frac{z^+_\alpha}{z^+_{\alpha'}};\tau\right)}{\eta(\tau)}\right]; \\
    \mathbb{Q}_{+-}^{-1}
    &=\left[\prod_{\alpha=1}^{n_+}\prod_{\beta=1}^{n_-}\frac{\theta_{11}\left(\frac{z^+_\alpha}{z^-_{\beta}};\tau\right)}{\eta(\tau)}\right]; \\
    \mathbb{Q}_{--}^{-1}
    &=\left[\prod_{n_-\geq\beta>\beta'\geq1}\frac{\theta_{11}\left(\frac{z^-_\beta}{z^-_{\beta'}};\tau\right)}{\eta(\tau)}\right],
\end{align}
\end{subequations}
and $\mathbb{Q}^{-1}$ is the theta function of $\widehat{A}_{n_+-1|n_--1}$. See \cite{kac1984infinite,kac1990infinite} for more details about the higher rank theta function.
$\mathbb{Q}$ is also thought of as the free fermionic correlation function on a torus. See for example, \cite{Fay:1973,Kajihara:2003IM} for a determinant formula.
%\paragraph{}

As mentioned earlier, in order to have non-vanishing interaction for the particles in different sectors, one may deliberately modify the metric in \eqref{metric} to: 
\begin{equation}\label{metric deform}
     \begin{pmatrix}
     +\mathbf{1}_{n_+} & 0 \\
    0 & -\theta\mathbf{1}_{n_-}
    \end{pmatrix}.
\end{equation}
The metric deformation in \eqref{metric deform} breaks $U(n_+|n_-)$ to $U(n_+)\times U(n_-)\subset U(n_+|n_-)$, which leads to the deformation of commutation relation for canonical momenta:
\begin{subequations}\label{commutation}
\begin{align}
    &[\rx_{\alpha}^+,p^+_{\alpha}]=-\eps\delta_{\alpha\alpha'};\quad \alpha,\alpha'=1,\dots,n_+, \\
    &[\rx_{\beta}^-,p^-_{\beta'}]=\theta\eps\delta_{\beta\beta'};\quad \beta,\beta'=1,\dots,n_-, \\
    &[\rx_{\alpha}^+,p^-_\beta]=0=[\rx_{\beta}^-,p^+_{\alpha}]; \quad \alpha=1,\dots,n_+;\quad\beta=1,\dots,n_-, 
\end{align}
\end{subequations}
This is equivalent to modifying $\Omega$-parameter in the negative sector by factor of $\theta$. 

To extract the resultant Hamiltonians, we first identify following property of $\mathbb{X}_\omega(x)$  \eqref{Xbb+} in the large $x$ limit:
\begin{align}
    {\mathbb{X}_{\omega}(x)}=
    \begin{cases}
    x\left[1+\frac{\eps}{x}+\frac{c_{1,\omega}}{x}+\frac{c_{2,\omega}}{x^2}+\cdots\right]\times {\mathbb{B}_\omega}, & \omega\in[\alpha];\\
    \left[x[1+\frac{\theta\eps}{x}+\frac{c_{1,\omega}}{x}+\frac{c_{2,\omega}}{x^2}+\cdots]\right]^{-1} \times {\mathbb{B}_\omega}, & \omega\in[\beta].
    \end{cases}
\end{align}
with $[\alpha]=\{0,\dots,n_+-1\}$ and $[\beta]=\{n_+,\dots,n_++n_--1\}$ based on the coloring function defined in \eqref{coloring}.
Such that the first and second Hamiltonians can be found by collecting the coefficients:
\begin{subequations}
\begin{align}
    h_1&=\sum_{\omega\in[\alpha]}c_{1,\omega}-\sum_{\omega\in[\beta]}c_{1,\omega}, \\
    -h_2&=\sum_{\omega\in[\alpha]}c_{2,\omega}-\sum_{\omega\in[\beta]}c_{2,\omega}.
\end{align}
\end{subequations}
We carefully set up the orbifolding such that $\mathbb{X}_{\omega}$ is of degree plus one when $\omega\in[\alpha]$, and of degree minus one when $\omega\in[\beta]$ by property of $Y_\omega(x)$ defined in \eqref{Y_w}.

To extract $\{c_{1,\omega},c_{2,\omega},\dots\}$, we consider the expansion in large $x$ limit. As stated, we are also required to re-scale $\Omega$-parameter in the negative sector according to the the new metric \eqref{metric deform}.  When $\omega\in[\alpha]$, large $x$ expansion of $\mathbb{X}_\omega(x)$ can be obtained as
\begin{align}
    &\frac{\mathbb{X}_\omega(x)}{\mathbb{B}_\omega} \nonumber\\
    &=x+\eps\nu_\omega-a^+_{c^{-1}(\omega+1)}+\eps+\frac{1}{x}\left[\frac{\eps^2}{2}\nu_\omega^2-\eps a^+_{c^{-1}(\omega+1)}\nu_\omega+\eps D_\omega\right. \nonumber \\
    &\left.\quad-m\sum_{\omega'\in[\alpha]}\left[\eps k\nabla^\fq_\omega+\left(\eps\nu_{\omega'}-a^+_{c^{-1}(\omega'+1)}\right)\nabla^\mathfrak{z}_{\omega'}\right]\log\mathbb{B}_\omega
    -\frac{m}{\theta}\sum_{\omega'\in[\beta]}\left[\theta\eps k\nabla^\fq_\omega+\left(\theta\eps\nu_{\omega'}+a^-_{c^{-1}(\omega'+1)}\right)\nabla^\mathfrak{z}_{\omega'}\right]\log\mathbb{B}_\omega\right]+\cdots.
\end{align}
As for $\omega\in[\beta]$, we have the large $x$ expansion of $\mathbb{X}_\omega(x)$ as
\begin{align}
    &\left[\frac{\mathbb{X}_\omega(x)}{\mathbb{B}_\omega}\right]^{-1} \nonumber\\ 
    &=x-\theta\eps\nu^-_\omega-a^-_{c^{-1}(\omega+1)}+\theta\eps+\frac{1}{x}\left[\frac{\theta^2\eps^2}{2}(\nu_\omega^-)^2+\theta \eps a^-_{c^{-1}(\omega+1)}\nu_\omega^--\theta \eps D_\omega+\right. \nonumber\\
    &\quad\left.+m\sum_{\omega'\in[\alpha]}\left[\eps k\nabla^\fq_\omega+\left( \eps \nu_{\omega'}-a^+_{c^{-1}(\omega'+1)}\right)\nabla^\mathfrak{z}_{\omega'}\right]\log\mathbb{B}_\omega
    +\frac{m}{\theta}\sum_{\omega'\in[\beta]}\left[\theta\eps k\nabla^\fq_\omega+\left( \theta \eps \nu_{\omega'}+a^-_{c^{-1}(\omega'+1)}\right)\nabla^\mathfrak{z}_{\omega'}\right]\log\mathbb{B}_\omega\right]+\cdots.
\end{align}
We can now take the total $\mathbb{X}(x)$ to be the following combination for normalizing the overall coefficient:
\begin{align}
    \mathbb{X}(x)\implies\prod_{\omega\in[\alpha]}\left[\frac{\mathbb{X}_{\omega}(x)}{\mathbb{B}_\omega}\right]\prod_{\omega\in[\beta]}\left[\theta\frac{\mathbb{X}_{\omega}(x)}{\mathbb{B}_\omega}\right].
\end{align}
The first Hamiltonian is found to be
\begin{align}
    h_1
    &=\sum_{\omega\in[\alpha]}c_{1,\omega}-\frac{1}{\theta}\sum_{\omega\in[\beta]}c_{1,\omega} \nonumber\\
    &=\sum_{\omega\in[\alpha]}\left[\eps\nu_\omega-a^+_{c^{-1}(\omega'+1)}\right]-\frac{1}{\theta}\sum_{\omega\in[\beta]}\left[-\theta\eps\nu_\omega-a^-_{c^{-1}(\omega+1)}\right] \nonumber \\
    &=\sum_{\alpha}p_\alpha^+-\frac{1}{\theta}\sum_{\beta}p_\beta^-.
\end{align}
We will rewrite conjugate momentum $\{p_\alpha^+,p_\beta^-\}$ with respect to the commutation relation \eqref{commutation}:
\begin{align}
    p_\alpha^+=\eps\frac{\partial}{\partial\rx_\alpha^+};\quad p_\beta^-=-\theta\eps\frac{\partial}{\partial\rx_\beta^-},
\end{align}
such that the the first momentum can be written as
\begin{align}
    h_1=\epsilon\left[\sum_{\alpha=1}^{n_+}\frac{\partial}{\partial\rx_\alpha^+}+\sum_{\beta=1}^{n_-}\frac{\partial}{\partial\rx_\beta^-}\right],
\end{align}
which agrees with the conserved charges constructed using Dunkl operator \cite{Chen:2019vvt}. The second Hamiltonian can be found by
\begin{align}\label{super h_2 deformed}
    -h_{2}
    =&\sum_{\omega\in[\alpha]}c_{2,\omega}-\frac{1}{\theta}\sum_{\omega\in[\beta]}c_{2,\omega} \nonumber\\
    =&\sum_{\alpha=1}^{n_+}\frac{(p^+_\alpha)^2}{2}-\frac{1}{\theta}\sum_{\beta=1}^{n_-}\frac{(p^-_\beta)^2}{2} 
    -\eps k\sum_{\omega\in[\alpha]}\left[\eps k\nabla^\fq_\omega+p_{c^{-1}(\omega+1)}\nabla^\mathfrak{z}_{\omega}\right](\log\mathbb{Q}_{++}-\frac{1}{\theta}\log\mathbb{Q}_{+-})\nonumber \\
    &-\eps k\sum_{\omega\in[\beta]}\frac{1}{\theta}\left[\eps k\nabla^\fq_\omega+p_{c^{-1}(\omega+1)}\nabla^\mathfrak{z}_{\omega}\right](\log\mathbb{Q}_{+-}-\frac{1}{\theta}\log\mathbb{Q}_{--}).
\end{align}
Following \cite{Chen:2019vvt}, the equation \eqref{super h_2 deformed} is equivalent to the following expression under canonical transformation and \eqref{Heat eq for Q}
\begin{align}\label{Deformed h_2}
    -\frac{h_{2}'}{\eps^2}
    =&\sum_{\alpha=1}^{n_+}\frac{(p^+_\alpha)^2}{2\eps^2}+\frac{k}{2}(\nabla^\mathfrak{z}_{c(\alpha)})^2\log\mathbb{Q}_{++}-\frac{k^2}{2}(\nabla^\mathfrak{z}_{c(\alpha)})^2\log\mathbb{Q}_{++} -\frac{k}{2\theta}(\nabla^\mathfrak{z}_{c(\alpha)})^2\log\mathbb{Q}_{+-}-\frac{k^2}{2\theta^2}(\nabla^\mathfrak{z}_{c(\alpha)})^2\log\mathbb{Q}_{+-} \nonumber \\
    &+\sum_{\beta=1}^{n_-}-\frac{1}{\theta}\frac{(p^-_\beta)^2}{2\eps^2}-\frac{k}{2\theta}(\nabla^\mathfrak{z}_{c(\beta)})^2\log\mathbb{Q}_{--}+\frac{k^2}{2\theta^3}(\nabla^\mathfrak{z}_{c(\beta)})^2\log\mathbb{Q}_{--} +\frac{k}{2}(\nabla^\mathfrak{z}_{c(\beta)})^2\log\mathbb{Q}_{+-}+\frac{k^2}{2\theta}(\nabla^\mathfrak{z}_{c(\beta)})^2\log\mathbb{Q}_{+-} \nonumber \\
    =&\sum_{\alpha=1}^{n_+}\frac{1}{2}\frac{\partial^2}{\partial(\rx_\alpha^+)^2}+\sum_{\beta=1}^{n_-}-\frac{\theta}{2}\frac{\partial^2}{\partial(\rx_\beta^-)^2} 
    - k(k-1)\sum_{n_+\geq\alpha>\alpha'\geq1}\wp(\rx_\alpha^+-\rx_{\alpha'}^+;\tau)-\left(\frac{k}{\theta}-\frac{k^2}{\theta^3}\right)\sum_{n_-\geq\beta>\beta'\geq1}\wp(\rx_\beta^--\rx_{\beta'}^-;\tau) \nonumber\\ &-\frac{1}{2}\left(\frac{k}{\theta}+\frac{k^2}{\theta^2}-k-\frac{k^2}{\theta}\right)\sum_{\alpha=1}^{n_+}\sum_{\beta=1}^{n_-}\wp(\rx_\alpha^+-\rx_\beta^-;\tau).
\end{align}
Taking the undeformed limit $\theta=1$, which has vanishing interacting potential between the two sectors as we expected for $\gamma\in R_{+-}$ discussed earlier. And when $\theta=-1$, the system becomes ordinary eCM with $n_++n_-$ particles, which agrees with the definition of deformed metric \eqref{metric deform}.
The last special value for deformation parameter is when $\theta=k$, for which deformed Hamiltonian \eqref{Deformed h_2} reads
\begin{align}\label{theta=k deformed h_2}
    \frac{h_2'}{\eps^2}
    =&-\sum_{\alpha=1}^{n_+}\frac{1}{2}\frac{\partial^2}{\partial(\rx_\alpha^+)^2}+\sum_{\beta=1}^{n_-}+\frac{k}{2}\frac{\partial^2}{\partial(\rx_\beta^-)^2} +k(k-1)\sum_{n_+\geq\alpha>\alpha'\geq1}\wp(\rx_\alpha^+-\rx_{\alpha'}^+;\tau) \nonumber \\
    &+\left(1-\frac{1}{k}\right)\sum_{n_-\geq\beta>\beta'\geq1}\wp(\rx_\beta^--\rx_{\beta'}^-;\tau) +\left(1-k\right)\sum_{\alpha=1}^{n_+}\sum_{\beta=1}^{n_-}\wp(\rx_\alpha^+-\rx_\beta^-;\tau).
\end{align}
We have recovered Hamiltonian of edCM in \eqref{Hamiltonian edCM}.

\subsection{Toda Limit}
%\paragraph{}

Finally we would like to consider the Toda limit $m\to\infty$, which is equivalent to have $k=\frac{m+\eps}{\eps}\approx\frac{m}{\eps}\gg1$. We first need to consider trigonometric limit $\fq\to0$ such that the combination $\fq m^{2n_+-2n_-}$ is kept finite. The Weierstrass $\wp$-unction has the following trigonometric expansion
\begin{align}
    \wp(x)=\frac{1}{4}\frac{1}{\sinh^2(\frac{x}{2})}-2\sum_{n=1}^\infty\frac{n\fq^n}{1-\fq^n}\cosh(nx).
\end{align}
Under the limit $\fq\to0$, all the higher terms are suppressed. Unlike the edCM system, the interaction of Toda lattice does not depend on the norm of the root system. 
To reduce from edCM to super Toda lattice, we also need to remove the contribution from the potential which is proportional to the norm of root system.
The Hamiltonian in \eqref{Deformed h_2} now becomes:
\begin{align}
    -h_2'=&\sum_{\alpha}\frac{(p^+_\alpha)^2}{2}-\frac{1}{\theta}\sum_{\beta}\frac{(p^-_\beta)^2}{2} \nonumber \\
    &-\sum_{\alpha>\alpha'}\frac{m^2}{4\sinh^2\frac{(\rx_\alpha^+-\rx_{\alpha'}^+)}{2}}+\sum_{\beta>\beta'}\frac{m^2/\theta^4}{4\sinh^2\frac{(\rx_\beta^--\rx_{\beta'}^-)}{2}}-\sum_{\alpha,\beta}\frac{m^2/\theta^2}{4\sinh^2\frac{(\rx_\alpha^+-\rx_\beta^-)}{2}}.
\end{align}
We will now turn off the deformation by setting $\theta=1$.%
\footnote{%
Although we have tuned $\theta = 1$ to reproduce the super Toda lattice system, it would be interesting to keep the deformation parameter $\theta$, and study the corresponding $\theta$-deformation of super Toda system.}
and redefine the coordinate system by
\begin{align}
    e^{\mathfrak{x}_{\omega}-\mathfrak{x}_{\omega-1}}\to\Lambda^{2}e^{ (\mathfrak{x}_{\omega}-\mathfrak{x}_{\omega-1})}, 
\end{align}
such that $\Lambda^{2n_+-2n_-}=\fq$. For $n_+>n_-$
We now further set $\Lambda^2\to0$ with $m^2\Lambda^2$ fixed. For $n_+<n_-$, we further set $\Lambda^2\to\infty$ such that $\frac{m^2}{\Lambda^2}$ fixed. In both ways
the only surviving terms are defined on fundamental root, as we have discussed in Section \ref{Special Dynkin}.

\section{Discussions and Future Directions}
%\paragraph{}

Let us end with discussing about the few loose ends in this note and commenting on few future directions.
\begin{itemize}
    \item It would be desirable to give an explicit realization of the deformation parameter $\theta$ in \eqref{inner product} and \eqref{metric} via gauge theoretical construction. In constructing the relevant quantum integrable system, it is implemented at the level of superalgebra in order to have non-vanishing interaction between two otherwise independent CM systems. However the gauge theoretical interpretation of such a parameter remains rather unclear, and we merely regarded it as the rescaling of $\Omega$-background parameter $\epsilon$.  We can consider realizing such a 
    $\theta$-deformation through the intersecting D-brane configuration at an angle \cite{Mikhaylov:2014aoa,Rapcak:2019wzw}, in particular in the gauge origami setup for the folded instanton configuration \cite{Chen:2019vvt}, $\theta$-deformation parameter naturally equals to $k$ by having more $\epsilon$-parameters.
    
    \item The edCM system can also be constructed from gauge origami setup for gauge theories with ordinary gauge group \cite{Nekrasov:2017gzb,Chen:2019vvt}. By applying similar gauge origami construction to the super gauge group, it will be interesting to construct a possible double version of elliptic double Calogero-Moser (d$^2$CM) system. The resultant quantum integrable system should take value on $\mathbb{Z}_2 \times \mathbb{Z}_2$-graded superalgebra~\cite{Rittenberg:1978mr,Rittenberg:1978df}.  
    
    \item For theories with ordinary gauge group, the wave functions of the associated quantum integrable systems are identified with the orbifolded instanton partitions up to overall abelian factors \cite{Nekrasov:2017gzb}. It is natural to ask if the same results also holds in general for their super gauge group counterparts.
    \item It would also be interesting to understand more about the various co-dimension two surface defects within the supergroup gauge theories, their effective world volume theories and its class-$\mathcal{S}$ interpretation~\cite{Alday:2009fs}.
    
    \item Despite name resemblance, super spin chain system of \eqref{Bethe} is different from so called superalgebra spin system discussed in various papers~\cite{Evans:1996bu,Beisert:2003yb}, which the spin system is defined on, e.g., $\mathfrak{sl}(n_+|n_-)$ algebra. 
    Actually the spin symmetry depends on the quiver structure under the Bethe/Gauge correspondence~\cite{Nekrasov:2012xe,Nekrasov:2013xda}, and such quiver gauge theories corresponding to $\mathfrak{sl}(n_+|n_-)$ superalgebra spin chains have been constructed in~\cite{Orlando:2010uu,Nekrasov:2018gne,Zenkevich:2018fzl}. 
    \item In this paper, we have focused on 4d $\mathcal{N}=2$ supergroup gauge theory to study its correspondence to the rational-type integrable system.
    We could generalize this setup to 5d $\mathcal{N}=1$ and 6d $\mathcal{N}=(1,0)$ theories with supergroup gauge symmetry to explore the correspondence to the trigonometric and elliptic integrable systems~\cite{Nekrasov:1996cz}.
\end{itemize}

\subsection*{Acknowledgements}
The work of HYC was supported in part by Ministry of Science and Technology (MOST)
through the grant 107 -2112-M-002-008-.
The work of TK was supported in part by the French ``Investissements d'Avenir'' program, project ISITE-BFC (No.~ANR-15-IDEX-0003), JSPS Grant-in-Aid for Scientific Research (No.~JP17K18090), the MEXT-Supported Program for the Strategic Research Foundation at Private Universities ``Topological Science'' (No.~S1511006), JSPS Grant-in-Aid for Scientific Research on Innovative Areas ``Topological Materials Science'' (No.~JP15H05855), and ``Discrete Geometric Analysis for Materials Design'' (No.~JP17H06462). The work of NL is supported by Simons Center for Geometry and Physics and State of New York.

\newpage
\appendix
\section{List of Relevant Mathematical Functions}
%\paragraph{}

In this appendix, we provide some relevant details about the mathematical functions used in the main text.

\subsection{Random Partition}
%\paragraph{}

A partition is defined as a way of expressing a non-negative integer $n$ as a summation over other non-negative integers. Each partition can be labeled by a Young diagram $\lambda=(\lambda_1 \ge \lambda_2 \ge \dots \ge \lambda_{\ell(\lambda)} > 0)$ with $\lambda_i\in\mathbb{N}$ such that 
\begin{equation}
n=|\lambda|=\sum_{i=1}^{\ell(\lambda)}\lambda_i,
\end{equation} 
where $\ell(\lambda)$ denotes the number of rows in $\lambda$.
We define the generating function of such a partition as
\begin{subequations}\label{phi}
\begin{align}
&\sum_{\lambda}\mathfrak{q}^{|\lambda|}=\frac{1}{(\mathfrak{q};\mathfrak{q})_\infty},\quad(\mathfrak{q};\mathfrak{q})_\infty=\prod_{n=1}^\infty\left(1-\mathfrak{q}^n\right); \\
&\sum_{\lambda}t^{\ell(\lambda)}\mathfrak{q}^{|\lambda|}=\frac{1}{(\mathfrak{q}t;\mathfrak{q})_\infty};\quad (\mathfrak{q}t;\mathfrak{q})_\infty=\prod_{n=1}^\infty(1-t\mathfrak{q}^n).
\end{align}
\end{subequations}
Here we also define the $\mathfrak{q}$-shifted factorial (the $\mathfrak{q}$-Pochhammer symbol) as
\begin{equation}\label{q-Pochhammer}
(z;q)_n=\prod_{m=0}^{n-1}(1-zq^m).
\end{equation}

\subsection{The Elliptic Function}
%\paragraph{}

Here we fix our notation for the elliptic functions. The so-called Dedekind eta function is denoted as
\begin{equation}\label{eta}
\eta(\tau)=e^{\frac{\pi i\tau}{12}}(\mathfrak{q};\mathfrak{q})_\infty,
\end{equation}
where $\mathfrak{q}=\exp \left( 2 \pi i \tau \right)$.
The first Jacobi theta function is denoted as:
\begin{equation}\label{theta}
\theta_{11}(z;\tau)=ie^{\frac{\pi i\tau}{4}}z^{\frac{1}{2}}(\mathfrak{q};\mathfrak{q})_\infty (\mathfrak{q}z;\mathfrak{q})_\infty(z^{-1};\mathfrak{q})_\infty,
\end{equation} 
whose series expansion 
\begin{equation}\label{theta2}
\theta_{11}(z;\tau)=i\sum_{r\in\mathbb{Z}+\frac{1}{2}}(-1)^{r-\frac{1}{2}}z^re^{\pi i\tau r^2}=i\sum_{r\in\mathbb{Z}+\frac{1}{2}}(-1)^{r-\frac{1}{2}}e^{rx}e^{\pi i\tau r^2},
\end{equation}
implies that it obeys the heat equation
\begin{equation}
\frac{1}{\pi i}\frac{\partial}{\partial\tau}\theta_{11}(z;\tau)=(z\partial_z)^2\theta_{11}(z;\tau).
\end{equation}
The Weierstrass $\wp$-function
\be\label{Def:p-function}
\wp(z) = \frac{1}{z^2}+\sum_{p, q \ge 0} \left\{\frac{1}{(z+p+q\tau)^2}-\frac{1}{(p+q\tau)^2}\right\},
\ee 
is related to theta and eta functions by
\begin{equation}
\wp(z;\tau)=-(z\partial_z)^2\log\theta_{11}(z;\tau)+\frac{1}{\pi i}\partial_\tau\log\eta(\tau).
\end{equation}

\subsection{Higher rank Theta function}
Let us also define
\begin{equation}
\Theta_{\widehat{A}_{N-1}}(\vec{z};\tau)=\eta(\tau)^{N}\prod_{\alpha>\beta}\frac{\theta_{11}(z_\alpha/z_\beta;\tau)}{\eta(\tau)}
\end{equation}
as the rank $N-1$ theta function, which also satisfies the generalized heat equation
\begin{equation}\label{heat}
N\frac{\partial}{\partial\tau}\Theta_{\widehat{A}_{N-1}}(\vec{z};\tau)=\pi i\Delta_{\vec{z}}\Theta_{\widehat{A}_{N-1}}(\vec{z};\tau),
\end{equation}
with the $N$-variable Laplacian:
\begin{align}
    \Delta_{\vec{z}}=\sum_{\omega=0}^{N-1}(z_\omega\partial_{z_\omega})^2.
\end{align}

\subsection{The Orbifolded Partition}
%\paragraph{}

For the purpose in the main text, we consider the orbifolded coupling 
\begin{align}
    \mathfrak{q}=\prod_{\omega=0}^{N-1}\mathfrak{q}_\omega;\quad \mathfrak{q}_{\omega+N}=\mathfrak{q}_\omega,
\end{align}
and
\begin{align}
    \mathfrak{q}_\omega=\frac{z_\omega}{z_{\omega-1}};\quad z_{\omega+N}=\mathfrak{q}z_{\omega}.
\end{align}
We also consider the orbifolded version of the generating function of partitions $(\mathfrak{q};\mathfrak{q})_\infty^{-1}$ in \eqref{phi}. Given a finite partition $\lambda=(\lambda_1,\dots,\lambda_{\ell(\lambda)})$, we define
\begin{equation}
\mathbb{Q}^{\lambda}_\omega
=\prod_{j=1}^{\lambda_1}\mathfrak{q}_{\omega+1-j}^{\lambda_j^\text{T}}
=\prod_{i=1}^{\ell(\lambda)}\frac{z_\omega}{z_{\omega-\lambda_{i}}}.
\end{equation}
where the transposed partition is denoted by $\lambda^\text{T}$.
The summation over all possible partitions is given by
\begin{equation}\label{Q-form}
\mathbb{Q}_\omega=\sum_{\lambda}\mathbb{Q}^{\lambda}_\omega=\sum_{\lambda}\prod_{i=1}^{\ell(\lambda)}\left(\frac{z_\omega}{z_{\omega-\lambda_{i}}}\right)=\sum_{l_0,\dots,l_{N-1},l\geq0}\prod_{\alpha=1}^{N-1}\left(\frac{z_\omega}{z_{\alpha}}\right)^{l_\alpha}\mathfrak{q}^l.
\end{equation}
%%%%%%%
The function $\mathbb{Q}(\vec{z};\tau)$ is the orbifolded version of the generating function of partitions \eqref{phi}, 
\begin{align}\label{Q}
\mathbb{Q}(\vec{z};\tau)
&=\prod_{\omega=0}^{N-1}\mathbb{Q}_\omega(\vec{z};\tau) \nonumber\\
&=\prod_{N-1 \geq \alpha>\beta \geq 0}\frac{1}{(\frac{z_\alpha}{z_\beta};\mathfrak{q})_\infty(\mathfrak{q}\frac{z_\beta}{z_\alpha};\mathfrak{q})_\infty}\prod_{\alpha=0}^{N-1}\frac{1}{(\mathfrak{q};\mathfrak{q})_\infty} \nonumber\\
&=\prod_{N-1\geq\alpha>\beta\geq0}\frac{\mathfrak{q}^{1/12}\eta(\tau)\sqrt{z_\alpha/z_\beta}}{\theta_{11}(z_\alpha/z_\beta;\tau)}\times\left[\frac{\mathfrak{q}^{1/24}}{\eta(\tau)}\right]^N \nonumber\\
&=\left[\eta(\tau)^{-N}\prod_{N-1\geq\alpha>\beta\geq0}\frac{\eta(\tau)}{\theta_{11}(z_\alpha/z_\beta;\tau)}\right]\frac{\mathfrak{q}^{N^2/24}}{\vec{z}^{\vec{\rho}}} \nonumber\\
&=\frac{1}{\Theta_{\widehat{A}_{N-1}}(\vec{z};\tau)}\frac{\mathfrak{q}^{N^2/24}}{\vec{z}^{\vec{\rho}}},
\end{align}
where $\vec{\rho}$ is the Weyl vector of $SU(N)$ Lie group, whose entries are given as
\begin{equation}
\vec{\rho}=(\rho_0,\dots,\rho_{N-1});\quad\rho_\omega=\omega-\frac{N-1}{2};\quad |\vec{\rho}|^2=\sum_{\omega=0}^{N-1}\rho_\omega^2=\frac{N(N^2-1)}{12};\quad\vec{z}^{\vec{\rho}}=\prod_{\omega=0}^{N-1}z_\omega^{\rho_\omega}.
\end{equation}
Using eq.~\eqref{heat}, it is easy to prove that the $\mathbb{Q}$-function satisfies
\begin{equation}\label{Heat eq for Q}
0=\sum_{\omega}\nabla^{\mathfrak{q}}_\omega\log\mathbb{Q}-\frac{1}{2}\Delta_{\vec{z}}\log\mathbb{Q}+\frac{1}{2}\sum_\omega(\nabla^z_\omega\log\mathbb{Q})^2,
\end{equation}
with 
\begin{equation}
\sum_\omega\nabla_\omega^\mathfrak{q}=N\nabla^\mathfrak{q}+\vec{\rho}\cdot{\nabla}^{\vec{z}} .
\end{equation}

\section{Superalgebra and Supermatrix}\label{Superalgebra}
In this short appendix we also specify our conventions of superalgebra and supermatrices used in the main text.
See also \cite{Kac:1977em,Quella:2013oda} for details.
\subsection{Superalgebra}
%\paragraph{}

A superalgebra is a $\bbZ_2$-graded algebra. It is an algebra over a commutative ring or field $K$ with decomposition into even and odd elements and a multiplication operator that respects the grading. A superalgebra over $K$ is defined by direct sum decomposition
\begin{align}
    A=A_0\oplus A_1
\end{align}
with a bilinear multiplication $A\times A\to A$ such that 
\begin{align}
    A_i\times A_j\subseteq A_{i+j}.
\end{align}
A parity is assigned to every element $x\in A$, denoted as $|x|$, is either $0$ or $1$ depending on whether $x$ is in $A_0$ or $A_1$. Define supercommutator by
\begin{align}\label{super odd}
    [x,y]=xy-(-1)^{|x||y|}yx.
\end{align}
A superalgebra $A$ is said to be commutative if 
\begin{align}
    [x,y]=0\quad \forall x,y\in A.
\end{align}

\subsection{Supermatrix}
%\paragraph{}

A supermatrix is a $\bbZ_2$-graded analog of ordinary matrix. Specially it is a $2\times2$ block matrix with entries in superalgebra $R$. $R$ can be either commutative superalgebra (e.g. Grassmannian algebra) or ordinary field. A supermatrix of dimension $(r|s)\times(p|q)$ is a matrix
\begin{align}\label{supermatrix blockform}
    X=\begin{pmatrix}
    X_{00} & X_{01} \\
    X_{10} & X_{11}
    \end{pmatrix}
\end{align}
with $r+s$ rows and $p+q$ columns. The block matrices $X_{00}$ and $X_{11}$ consist solely of even graded element in $R$, while
$X_{01}$ and $X_{10}$ consist solely of odd graded elements in $R$. 
If $X$ is a square matrix, its supertrace is defined as
\begin{align}
    \operatorname{str}(X)=\operatorname{tr}(X_{00})-\operatorname{tr}(X_{11}).
\end{align}
For an invertible supermatrix over commutative super algebra, its superdeterminant (Berezinian) is defined as
\begin{align}
    \operatorname{sdet} X =\det(X_{00}-X_{01}X^{-1}_{11}X_{10})\det(X_{11})^{-1} = \det (X_{00}) \det(X_{11} - X_{10}X_{00}^{-1}X_{01})^{-1}.
\end{align}

%\newpage
%%%%%% Bibliography %%%%%%
\bibliographystyle{utphys}
\bibliography{Super}

\providecommand{\href}[2]{#2}\begingroup\raggedright\begin{thebibliography}{10}

\bibitem{Okuda:2006fb}
T.~Okuda and T.~Takayanagi, ``{Ghost D-branes},''
  \href{http://dx.doi.org/10.1088/1126-6708/2006/03/062}{{\em JHEP} {\bfseries
  03} (2006) 062},
\href{http://arxiv.org/abs/hep-th/0601024}{{\ttfamily arXiv:hep-th/0601024
  [hep-th]}}.
%%CITATION = HEP-TH/0601024;%%.

\bibitem{Dijkgraaf:2016lym}
R.~Dijkgraaf, B.~Heidenreich, P.~Jefferson, and C.~Vafa, ``{Negative Branes,
  Supergroups and the Signature of Spacetime},''
  \href{http://dx.doi.org/10.1007/JHEP02(2018)050}{{\em JHEP} {\bfseries 02}
  (2018) 050},
\href{http://arxiv.org/abs/1603.05665}{{\ttfamily arXiv:1603.05665 [hep-th]}}.
%%CITATION = ARXIV:1603.05665;%%.

\bibitem{Nekrasov:2018xsb}
N.~Nekrasov and N.~Piazzalunga, ``{Magnificent Four with Colors},''
  \href{http://dx.doi.org/10.1007/s00220-019-03426-3}{{\em Commun. Math. Phys.}
  {\bfseries 372} no.~2, (2019) 573--597},
\href{http://arxiv.org/abs/1808.05206}{{\ttfamily arXiv:1808.05206 [hep-th]}}.
%%CITATION = ARXIV:1808.05206;%%.

\bibitem{Vafa:2001qf}
C.~Vafa, ``{Brane / anti-brane systems and U($N|M$) supergroup},''
\href{http://arxiv.org/abs/hep-th/0101218}{{\ttfamily arXiv:hep-th/0101218
  [hep-th]}}.
%%CITATION = HEP-TH/0101218;%%.

\bibitem{Kimura:2019msw}
T.~Kimura and V.~Pestun, ``{Super instanton counting and localization},''
\href{http://arxiv.org/abs/1905.01513}{{\ttfamily arXiv:1905.01513 [hep-th]}}.
%%CITATION = ARXIV:1905.01513;%%.

\bibitem{Martinec:1995by}
E.~J. Martinec and N.~P. Warner, ``{Integrable systems and supersymmetric gauge
  theory},'' \href{http://dx.doi.org/10.1016/0550-3213(95)00588-9}{{\em Nucl.
  Phys.} {\bfseries B459} (1996) 97--112},
\href{http://arxiv.org/abs/hep-th/9509161}{{\ttfamily arXiv:hep-th/9509161
  [hep-th]}}.
%%CITATION = HEP-TH/9509161;%%.

\bibitem{Gorsky:1994dj}
A.~Gorsky and N.~Nekrasov, ``{Elliptic Calogero-Moser system from
  two-dimensional current algebra},''
\href{http://arxiv.org/abs/hep-th/9401021}{{\ttfamily arXiv:hep-th/9401021
  [hep-th]}}.
%%CITATION = HEP-TH/9401021;%%.

\bibitem{Donagi:1995cf}
R.~Donagi and E.~Witten, ``{Supersymmetric Yang-Mills theory and integrable
  systems},'' \href{http://dx.doi.org/10.1016/0550-3213(95)00609-5}{{\em Nucl.
  Phys.} {\bfseries B460} (1996) 299--334},
\href{http://arxiv.org/abs/hep-th/9510101}{{\ttfamily arXiv:hep-th/9510101
  [hep-th]}}.
%%CITATION = HEP-TH/9510101;%%.

\bibitem{DHoker:1997hut}
E.~D'Hoker and D.~H. Phong, ``{Calogero-Moser systems in SU($N$) Seiberg-Witten
  theory},'' \href{http://dx.doi.org/10.1016/S0550-3213(97)00763-3}{{\em Nucl.
  Phys.} {\bfseries B513} (1998) 405--444},
\href{http://arxiv.org/abs/hep-th/9709053}{{\ttfamily arXiv:hep-th/9709053
  [hep-th]}}.
%%CITATION = HEP-TH/9709053;%%.

\bibitem{van1994super}
E.~van~der Lende, ``{Super-Toda lattices},''
  \href{http://dx.doi.org/10.1063/1.530586}{{\em J. Math. Phys.} {\bfseries 35}
  no.~3, (1994) 1233--1251}.

\bibitem{Evans:1996bu}
J.~M. Evans and J.~O. Madsen, ``{Dynkin diagrams and integrable models based on
  Lie superalgebras},''
  \href{http://dx.doi.org/10.1016/S0550-3213(97)00381-7}{{\em Nucl. Phys.}
  {\bfseries B503} (1997) 715--746},
\href{http://arxiv.org/abs/hep-th/9703065}{{\ttfamily arXiv:hep-th/9703065
  [hep-th]}}.
%%CITATION = HEP-TH/9703065;%%.

\bibitem{Beisert:2003yb}
N.~Beisert and M.~Staudacher, ``{The $\mathcal{N}=4$ SYM integrable super spin
  chain},'' \href{http://dx.doi.org/10.1016/j.nuclphysb.2003.08.015}{{\em Nucl.
  Phys.} {\bfseries B670} (2003) 439--463},
\href{http://arxiv.org/abs/hep-th/0307042}{{\ttfamily arXiv:hep-th/0307042
  [hep-th]}}.
%%CITATION = HEP-TH/0307042;%%.

\bibitem{HiJack}
F.~Atai, M.~Halln\"as, and E.~Langmann, ``{Orthogonality of super-Jack
  polynomials and a Hilbert space interpretation of deformed
  Calogero--Moser--Sutherland operators},''
  \href{http://dx.doi.org/10.1112/blms.12234}{{\em Bulletin London Math. Soc.}
  {\bfseries 51} no.~2, (Feb, 2019) 353--370},
  \href{http://arxiv.org/abs/1802.02016}{{\ttfamily arXiv:1802.02016
  [math.QA]}}.

\bibitem{sergeev2001superanalogs}
A.~Sergeev, ``{Superanalogs of the Calogero operators and Jack polynomials},''
  \href{http://dx.doi.org/10.2991/jnmp.2001.8.1.7}{{\em J. Nonlin. Math. Phys.}
  {\bfseries 8} no.~1, (2001) 59--64},
  \href{http://arxiv.org/abs/math/0106222}{{\ttfamily arXiv:math/0106222
  [math.RT]}}.

\bibitem{Chen:2019vvt}
H.-Y. Chen, T.~Kimura, and N.~Lee, ``{Quantum Elliptic Calogero-Moser Systems
  from Gauge Origami},'' \href{http://dx.doi.org/10.1007/JHEP02(2020)108}{{\em
  JHEP} {\bfseries 02} (2020) 108},
\href{http://arxiv.org/abs/1908.04928}{{\ttfamily arXiv:1908.04928 [hep-th]}}.
%%CITATION = ARXIV:1908.04928;%%.

\bibitem{Nekrasov:2017gzb}
N.~Nekrasov, ``{BPS/CFT correspondence V: BPZ and KZ equations from
  qq-characters},''
\href{http://arxiv.org/abs/1711.11582}{{\ttfamily arXiv:1711.11582 [hep-th]}}.
%%CITATION = ARXIV:1711.11582;%%.

\bibitem{Nikita-Shatashvili}
N.~A. Nekrasov and S.~L. Shatashvili,
  \href{http://dx.doi.org/10.1142/9789814304634_0015}{``{Quantization of
  Integrable Systems and Four Dimensional Gauge Theories},''} in {\em {XVIth
  International Congress on Mathematical Physics}}, pp.~265--289.
\newblock 2009.
\newblock
\href{http://arxiv.org/abs/0908.4052}{{\ttfamily arXiv:0908.4052 [hep-th]}}.
\newblock
%%CITATION = 0908.4052;%%.

\bibitem{HYC:2011}
H.-Y. Chen, N.~Dorey, T.~J. Hollowood, and S.~Lee, ``{A New 2d/4d Duality via
  Integrability},'' \href{http://dx.doi.org/10.1007/JHEP09(2011)040}{{\em JHEP}
  {\bfseries 09} (2011) 040},
\href{http://arxiv.org/abs/1104.3021}{{\ttfamily arXiv:1104.3021 [hep-th]}}.
%%CITATION = ARXIV:1104.3021;%%.

\bibitem{Dorey:2011pa}
N.~Dorey, S.~Lee, and T.~J. Hollowood, ``{Quantization of Integrable Systems
  and a 2d/4d Duality},'' \href{http://dx.doi.org/10.1007/JHEP10(2011)077}{{\em
  JHEP} {\bfseries 10} (2011) 077},
\href{http://arxiv.org/abs/1103.5726}{{\ttfamily arXiv:1103.5726 [hep-th]}}.
%%CITATION = ARXIV:1103.5726;%%.

\bibitem{Nekrasov:2012xe}
N.~Nekrasov and V.~Pestun, ``{Seiberg-Witten geometry of four dimensional $N=2$
  quiver gauge theories},''
\href{http://arxiv.org/abs/1211.2240}{{\ttfamily arXiv:1211.2240 [hep-th]}}.
%%CITATION = ARXIV:1211.2240;%%.

\bibitem{Kanno:2011fw}
H.~Kanno and Y.~Tachikawa, ``{Instanton counting with a surface operator and
  the chain-saw quiver},''
  \href{http://dx.doi.org/10.1007/JHEP06(2011)119}{{\em JHEP} {\bfseries 06}
  (2011) 119},
\href{http://arxiv.org/abs/1105.0357}{{\ttfamily arXiv:1105.0357 [hep-th]}}.
%%CITATION = ARXIV:1105.0357;%%.

\bibitem{Nakajima:2011yq}
H.~Nakajima, ``{Handsaw quiver varieties and finite W-algebras},'' {\em Moscow
  Math. J.} {\bfseries 12} (2012) 633,
\href{http://arxiv.org/abs/1107.5073}{{\ttfamily arXiv:1107.5073 [math.QA]}}.
%%CITATION = ARXIV:1107.5073;%%.

\bibitem{Kapustin:2009cd}
A.~Kapustin and N.~Saulina, ``{Chern-Simons-Rozansky-Witten topological field
  theory},'' \href{http://dx.doi.org/10.1016/j.nuclphysb.2009.07.006}{{\em
  Nucl. Phys.} {\bfseries B823} (2009) 403--427},
\href{http://arxiv.org/abs/0904.1447}{{\ttfamily arXiv:0904.1447 [hep-th]}}.
%%CITATION = ARXIV:0904.1447;%%.

\bibitem{Witten:2011zz}
E.~Witten, ``{Fivebranes and Knots},''
  \href{http://dx.doi.org/10.4171/QT/26}{{\em Quantum Topology} {\bfseries 3}
  (2012) 1--137},
\href{http://arxiv.org/abs/1101.3216}{{\ttfamily arXiv:1101.3216 [hep-th]}}.
%%CITATION = ARXIV:1101.3216;%%.

\bibitem{Mikhaylov:2014aoa}
V.~Mikhaylov and E.~Witten, ``{Branes And Supergroups},''
  \href{http://dx.doi.org/10.1007/s00220-015-2449-y}{{\em Commun. Math. Phys.}
  {\bfseries 340} no.~2, (2015) 699--832},
\href{http://arxiv.org/abs/1410.1175}{{\ttfamily arXiv:1410.1175 [hep-th]}}.
%%CITATION = ARXIV:1410.1175;%%.

\bibitem{Mikhaylov:2015nsa}
V.~Mikhaylov, ``{Analytic Torsion, 3d Mirror Symmetry And Supergroup
  Chern-Simons Theories},''
\href{http://arxiv.org/abs/1505.03130}{{\ttfamily arXiv:1505.03130 [hep-th]}}.
%%CITATION = ARXIV:1505.03130;%%.

\bibitem{Okazaki:2017sbc}
T.~Okazaki and D.~J. Smith, ``{Matrix supergroup Chern-Simons models for
  vortex-antivortex systems},''
  \href{http://dx.doi.org/10.1007/JHEP02(2018)119}{{\em JHEP} {\bfseries 02}
  (2018) 119},
\href{http://arxiv.org/abs/1712.01370}{{\ttfamily arXiv:1712.01370 [hep-th]}}.
%%CITATION = ARXIV:1712.01370;%%.

\bibitem{Aghaei:2018cbn}
N.~Aghaei, A.~M. Gainutdinov, M.~Pawelkiewicz, and V.~Schomerus,
  ``{Combinatorial Quantisation of $GL(1|1)$ Chern-Simons Theory I: The
  Torus},''
\href{http://arxiv.org/abs/1811.09123}{{\ttfamily arXiv:1811.09123 [hep-th]}}.
%%CITATION = ARXIV:1811.09123;%%.

\bibitem{osborn1981semiclassical}
H.~Osborn, ``Semiclassical functional integrals for self-dual gauge fields,''
\href{http://dx.doi.org/10.1016/0003-4916(81)90159-7}{{\em Annals of Physics}
  {\bfseries 135} no.~2, (1981) 373--415}.
%%CITATION = APNYA,135,373;%%.

\bibitem{Atiyah:1978ri}
M.~F. Atiyah, N.~J. Hitchin, V.~G. Drinfeld, and {\relax Yu}.~I. Manin,
  ``{Construction of Instantons},''
  \href{http://dx.doi.org/10.1016/0375-9601(78)90141-X}{{\em Phys. Lett.}
  {\bfseries A65} (1978) 185--187}.
[,133(1978)].
%%CITATION = PHLTA,A65,185;%%.

\bibitem{Nekrasov:2002qd}
N.~A. Nekrasov, ``{Seiberg-Witten prepotential from instanton counting},''
  \href{http://dx.doi.org/10.4310/ATMP.2003.v7.n5.a4}{{\em Adv. Theor. Math.
  Phys.} {\bfseries 7} no.~5, (2003) 831--864},
  \href{http://arxiv.org/abs/hep-th/0206161}{{\ttfamily arXiv:hep-th/0206161}}.

\bibitem{Nekrasov:2003rj}
N.~Nekrasov and A.~Okounkov, {\em {Seiberg-Witten theory and random
  partitions}}, vol.~244,
  \href{http://dx.doi.org/10.1007/0-8176-4467-9\_15}{pp.~525--596}.
\newblock 2006.
\newblock \href{http://arxiv.org/abs/hep-th/0306238}{{\ttfamily
  arXiv:hep-th/0306238}}.

\bibitem{Nekrasov:2015wsu}
N.~Nekrasov, ``{BPS/CFT correspondence: non-perturbative Dyson-Schwinger
  equations and qq-characters},''
  \href{http://dx.doi.org/10.1007/JHEP03(2016)181}{{\em JHEP} {\bfseries 03}
  (2016) 181},
\href{http://arxiv.org/abs/1512.05388}{{\ttfamily arXiv:1512.05388 [hep-th]}}.
%%CITATION = ARXIV:1512.05388;%%.

\bibitem{Frenkel:1998}
E.~Frenkel and N.~Reshetikhin,
  \href{http://dx.doi.org/10.1090/conm/248/03823}{``{The {$q$}-characters of
  representations of quantum affine algebras and deformations of
  {$\mathcal{W}$}-algebras},''} in {\em {Recent Developments in Quantum Affine
  Algebras and Related Topics}}, vol.~248 of {\em Contemp. Math.},
  pp.~163--205.
\newblock Amer. Math. Soc., 1999.
\newblock \href{http://arxiv.org/abs/math/9810055}{{\ttfamily math/9810055
  [math.QA]}}.

\bibitem{Knight:1995JA}
H.~Knight, ``{Spectra of Tensor Products of Finite Dimensional Representations
  of Yangians},'' \href{http://dx.doi.org/10.1006/jabr.1995.1123}{{\em J.
  Algebra} {\bfseries 174} (1995) 187--196}.

\bibitem{Kac:1977em}
V.~G. Kac, ``{Lie Superalgebras},''
\href{http://dx.doi.org/10.1016/0001-8708(77)90017-2}{{\em Adv. Math.}
  {\bfseries 26} (1977) 8--96}.
%%CITATION = ADMTA,26,8;%%.

\bibitem{Quella:2013oda}
T.~Quella and V.~Schomerus, ``{Superspace conformal field theory},''
  \href{http://dx.doi.org/10.1088/1751-8113/46/49/494010}{{\em J. Phys.}
  {\bfseries A46} (2013) 494010},
\href{http://arxiv.org/abs/1307.7724}{{\ttfamily arXiv:1307.7724 [hep-th]}}.
%%CITATION = ARXIV:1307.7724;%%.

\bibitem{Kimura:2020lmc}
T.~Kimura and Y.~Sugimoto, ``{Topological Vertex/anti-Vertex and Supergroup
  Gauge Theory},'' \href{http://dx.doi.org/10.1007/JHEP04(2020)081}{{\em JHEP}
  {\bfseries 04} (2020) 081},
\href{http://arxiv.org/abs/2001.05735}{{\ttfamily arXiv:2001.05735 [hep-th]}}.
%%CITATION = ARXIV:2001.05735;%%.

\bibitem{Orlando:2010uu}
D.~Orlando and S.~Reffert, ``{Relating Gauge Theories via Gauge/Bethe
  Correspondence},'' \href{http://dx.doi.org/10.1007/JHEP10(2010)071}{{\em
  JHEP} {\bfseries 10} (2010) 071},
\href{http://arxiv.org/abs/1005.4445}{{\ttfamily arXiv:1005.4445 [hep-th]}}.
%%CITATION = ARXIV:1005.4445;%%.

\bibitem{Zenkevich:2018fzl}
Y.~Zenkevich, ``{Higgsed network calculus},''
\href{http://arxiv.org/abs/1812.11961}{{\ttfamily arXiv:1812.11961 [hep-th]}}.
%%CITATION = ARXIV:1812.11961;%%.

\bibitem{Faddeev:1996iy}
L.~D. Faddeev, ``{How algebraic Bethe ansatz works for integrable model},'' in
  {\em {Relativistic gravitation and gravitational radiation. Proceedings,
  School of Physics, Les Houches, France, September 26-October 6, 1995}},
  pp.~149--219.
\newblock 1996.
\newblock
\href{http://arxiv.org/abs/hep-th/9605187}{{\ttfamily arXiv:hep-th/9605187
  [hep-th]}}.
\newblock
%%CITATION = HEP-TH/9605187;%%.

\bibitem{Mironov:2012uh}
A.~Mironov, A.~Morozov, Y.~Zenkevich, and A.~Zotov, ``{Spectral Duality in
  Integrable Systems from AGT Conjecture},''
  \href{http://dx.doi.org/10.1134/S0021364013010062}{{\em JETP Lett.}
  {\bfseries 97} (2013) 45--51},
  \href{http://arxiv.org/abs/1204.0913}{{\ttfamily arXiv:1204.0913 [hep-th]}}.
[Pisma Zh. Eksp. Teor. Fiz.97,49(2013)].
%%CITATION = ARXIV:1204.0913;%%.

\bibitem{Mironov:2012ba}
A.~Mironov, A.~Morozov, B.~Runov, Y.~Zenkevich, and A.~Zotov, ``{Spectral
  Duality Between Heisenberg Chain and Gaudin Model},''
  \href{http://dx.doi.org/10.1007/s11005-012-0595-0}{{\em Lett. Math. Phys.}
  {\bfseries 103} no.~3, (2013) 299--329},
\href{http://arxiv.org/abs/1206.6349}{{\ttfamily arXiv:1206.6349 [hep-th]}}.
%%CITATION = ARXIV:1206.6349;%%.

\bibitem{olshanetsky1983quantum}
M.~Olshanetsky and A.~Perelomov, ``{Quantum integrable systems related to Lie
  algebras},'' \href{http://dx.doi.org/10.1016/0370-1573(83)90018-2}{{\em
  Physics Reports} {\bfseries 94} no.~6, (1983) 313--404}.

\bibitem{Nekrasov:2013xda}
N.~Nekrasov, V.~Pestun, and S.~Shatashvili, ``{Quantum geometry and quiver
  gauge theories},'' \href{http://dx.doi.org/10.1007/s00220-017-3071-y}{{\em
  Commun. Math. Phys.} {\bfseries 357} no.~2, (2018) 519--567},
\href{http://arxiv.org/abs/1312.6689}{{\ttfamily arXiv:1312.6689 [hep-th]}}.
%%CITATION = ARXIV:1312.6689;%%.

\bibitem{Feigin:2017wnq}
B.~Feigin, M.~Jimbo, T.~Miwa, and E.~Mukhin, ``{Finite Type Modules and Bethe
  Ansatz for Quantum Toroidal ${\mathfrak{gl}_1}$},''
  \href{http://dx.doi.org/10.1007/s00220-017-2984-9}{{\em Commun. Math. Phys.}
  {\bfseries 356} no.~1, (2017) 285--327},
\href{http://arxiv.org/abs/1603.02765}{{\ttfamily arXiv:1603.02765 [math.QA]}}.
%%CITATION = ARXIV:1603.02765;%%.

\bibitem{Bourgine:2015szm}
J.-E. Bourgine, Y.~Mastuo, and H.~Zhang, ``{Holomorphic field realization of
  SH$^c$ and quantum geometry of quiver gauge theories},''
  \href{http://dx.doi.org/10.1007/JHEP04(2016)167}{{\em JHEP} {\bfseries 04}
  (2016) 167},
\href{http://arxiv.org/abs/1512.02492}{{\ttfamily arXiv:1512.02492 [hep-th]}}.
%%CITATION = ARXIV:1512.02492;%%.

\bibitem{Kimura:2015rgi}
T.~Kimura and V.~Pestun, ``{Quiver W-algebras},''
  \href{http://dx.doi.org/10.1007/s11005-018-1072-1}{{\em Lett. Math. Phys.}
  {\bfseries 108} no.~6, (2018) 1351--1381},
\href{http://arxiv.org/abs/1512.08533}{{\ttfamily arXiv:1512.08533 [hep-th]}}.
%%CITATION = ARXIV:1512.08533;%%.

\bibitem{kac1984infinite}
V.~G. Kac and D.~H. Peterson, ``{Infinite dimensional Lie algebras, theta
  functions and modular forms},''
\href{http://dx.doi.org/10.1016/0001-8708(84)90032-X}{{\em Adv. Math.}
  {\bfseries 53} (1984) 125--264}.
%%CITATION = ADMTA,53,125;%%.

\bibitem{kac1990infinite}
V.~G. Kac, {\em Infinite-dimensional Lie algebras}.
\newblock Cambridge University Press, 1990.

\bibitem{Fay:1973}
J.~D. Fay, \href{http://dx.doi.org/10.1007/BFb0060090}{{\em {Theta Functions on
  Riemann Surfaces}}}, vol.~352 of {\em Lecture Notes in Mathematics}.
\newblock Springer Berlin Heidelberg, 1973.

\bibitem{Kajihara:2003IM}
Y.~Kajihara and M.~Noumi, ``{Multiple elliptic hypergeometric series. An
  approach from the Cauchy determinant},''
  \href{http://dx.doi.org/10.1016/s0019-3577(03)90054-1}{{\em Indag. Mathem.}
  {\bfseries 14} (2003) 395--421},
  \href{http://arxiv.org/abs/math/0306219}{{\ttfamily arXiv:math/0306219
  [math.CA]}}.

\bibitem{Rapcak:2019wzw}
M.~Rap\v{c}\'ak, ``{On Extensions of $\widehat{\mathfrak{gl}(m|n)}$ Kac--Moody
  algebras and Calabi--Yau Singularities},''
  \href{http://dx.doi.org/10.1007/JHEP01(2020)042}{{\em JHEP} {\bfseries 01}
  (2020) 042},
\href{http://arxiv.org/abs/1910.00031}{{\ttfamily arXiv:1910.00031 [hep-th]}}.
%%CITATION = ARXIV:1910.00031;%%.

\bibitem{Rittenberg:1978mr}
V.~Rittenberg and D.~Wyler, ``{Generalized superalgebras},''
\href{http://dx.doi.org/10.1016/0550-3213(78)90186-4}{{\em Nucl. Phys.}
  {\bfseries B139} (1978) 189--202}.
%%CITATION = NUPHA,B139,189;%%.

\bibitem{Rittenberg:1978df}
V.~Rittenberg and D.~Wyler, ``{Sequences of $Z_2 \oplus Z_2$ graded lie
  algebras and superalgebras},''
\href{http://dx.doi.org/10.1063/1.523552}{{\em J. Math. Phys.} {\bfseries 19}
  (1978) 2193}.
%%CITATION = JMAPA,19,2193;%%.

\bibitem{Alday:2009fs}
L.~F. Alday, D.~Gaiotto, S.~Gukov, Y.~Tachikawa, and H.~Verlinde, ``{Loop and
  surface operators in N=2 gauge theory and Liouville modular geometry},''
  \href{http://dx.doi.org/10.1007/JHEP01(2010)113}{{\em JHEP} {\bfseries 01}
  (2010) 113},
\href{http://arxiv.org/abs/0909.0945}{{\ttfamily arXiv:0909.0945 [hep-th]}}.
%%CITATION = ARXIV:0909.0945;%%.

\bibitem{Nekrasov:2018gne}
N.~Nekrasov, ``{Superspin chains and supersymmetric gauge theories},''
  \href{http://dx.doi.org/10.1007/JHEP03(2019)102}{{\em JHEP} {\bfseries 03}
  (2019) 102},
\href{http://arxiv.org/abs/1811.04278}{{\ttfamily arXiv:1811.04278 [hep-th]}}.
%%CITATION = ARXIV:1811.04278;%%.

\bibitem{Nekrasov:1996cz}
N.~Nekrasov, ``{Five dimensional gauge theories and relativistic integrable
  systems},'' \href{http://dx.doi.org/10.1016/S0550-3213(98)00436-2}{{\em Nucl.
  Phys.} {\bfseries B531} (1998) 323--344},
\href{http://arxiv.org/abs/hep-th/9609219}{{\ttfamily arXiv:hep-th/9609219
  [hep-th]}}.
%%CITATION = HEP-TH/9609219;%%.

\end{thebibliography}\endgroup

\end{document}